\documentclass[aip,jcp,a4paper,reprint,floatfix,superscriptaddress]{revtex4-1}

\usepackage{graphicx}
\usepackage{color}
\usepackage{amsmath}
\usepackage{dsfont}
\usepackage{xspace}
\usepackage{bm}
\usepackage[caption=false]{subfig}

\usepackage{txfonts}

\definecolor{delftdark}{rgb}{0, .4, .6}

\usepackage[plainpages=false, pdfpagelabels,
    pdfpagemode=UseNone,
    pdfstartview=FitH,
    colorlinks=true,
    linkcolor=delftdark,
    citecolor=delftdark,
    a4paper=true,
    pdfstartpage=1,
    pdfauthor={Verzijl, Thijssen},
    pdfsubject={A DFT-based Molecular Transport Implementation in ADF/BAND},
    pdftitle={},
]{hyperref}

\newcommand{\itr}[1]{\ensuremath{^{(#1)}}}

\newcommand{\im}[1]{\ensuremath{\mathrm{Im}[#1]}}
\newcommand{\tr}[1]{\ensuremath{\mathrm{Tr}\{#1\}}}
\newcommand{\trr}[1]{\ensuremath{\mathrm{Tr}[#1]}}
\newcommand{\dphi}{\ensuremath{\Delta\phi} }

\newcommand{\pbc}{periodic boundary conditions\xspace}

\begin{document}

\title{A DFT-based Molecular Transport Implementation in ADF/BAND}

\author{C.J.O. Verzijl}\affiliation{Kavli Institute of Nanoscience, Delft University of Technology, 2628 CJ Delft, The Netherlands}\email{C.J.O.Verzijl@TUDelft.nl}
\author{J.M. Thijssen}\affiliation{Kavli Institute of Nanoscience, Delft University of Technology, 2628 CJ Delft, The Netherlands}

\date{\today}

\begin{abstract}
We present a novel implementation of the first-principles approach to molecular charge transport using the non-equilibrium Green's function formalism in combination with the ADF/\ BAND periodic band-structure DFT code, together with results for several example systems. As a proof of concept, we first discuss transport calculations on 1D chains of Li and Al atoms. We then present a detailed study of BDT and archetypal molecular wires from the OPE-family, sandwiched between 3D Au contacts, comparing well with results from the literature. Our implementation further allows us to make a comparison of 3D contacts with and without periodic boundary conditions, the latter being particularly useful for modeling the needle-shaped contacts used in break-junction experiments.
\end{abstract}

\maketitle

\section{Introduction}

In recent years, approaches to molecular transport based on density-functional theory (DFT) in combination with the non-equilibrium Green's function formalism (NEGF) have received considerable attention in the literature, driven by the rapid progress in experimental work on realizing single-molecule nanodevices.\cite{Agrait2003,Venkataraman2006,Osorio2007,Quek2007} A number of research codes,\cite{Taylor2001,Xue2002,Qian2007} as well as the TranSIESTA,\cite{Soler2002,Brandbyge2002,Stokbro2003a} TurboMole\cite{Ahlrichs1989,Evers2003,Rothig2006,Arnold2007} and SMEAGOL\cite{Rocha2006} production codes have been developed. The attractiveness of the approach is based on the strengths of DFT for treating realistic atomistic and molecular transport configurations self-consistently, starting from an \emph{ab initio} quantum chemical description, combined with an intuitive mapping to a Landauer-type expression for the conductance and current through the Green's function formalism.\cite{Meir1992,Datta2000}

The NEGF+DFT approach, already at the LDA level of theory, is known to work particularly well for transport in the strongly coupled regime\cite{Mozos2002,Stokbro2003} and through off-resonant transport levels.\cite{Li2008} It has, for example, been used for metallic wires and non-conjugated hydrocarbons (alkanes).\cite{Mozos2002,Li2008,Prasongkit2010,Zanolli2010} 

The strengths of DFT in this approach are, however, balanced by known weaknesses of the often-used approximate exchange-correlation potentials (\emph{e.g.}~LDA), which introduce self-interaction errors\cite{Perdew1981,Toher2005,Toher2007} and incorrect charging behavior due to the lack of a correct derivative discontinuity.\cite{Koentopp2006} The failure of common exchange correlation functionals to predict excited many-body states, as well as their mean-field character, hampers a proper handling of dynamic Coulomb correlations. This renders the method less suitable for weakly-coupled systems, particularly when one or more resonances are present inside the bias-window.

Nonetheless, despite the attractiveness of conceptually better-founded methods such as the GW approximation for dynamical response,\cite{Aryasetiawan1998,Thygesen2009} computational tractability has favored the popularity of the NEGF+DFT approach, especially when combined with better functionals from \emph{e.g.}~the GGA family. The approach has, for example, led to a better understanding of charge transport in thiolated phenyl systems\cite{Stokbro2003,Xue2003a,Xue2003b,Arnold2007} and single-molecule magnets,\cite{Barraza-Lopez2007,Park2010} among other systems of interest. 

Recent work on \emph{e.g.} self-interaction corrections,\cite{Toher2005,Toher2007} accounting for lead-renormalization and dielectric effects\cite{Neaton2006,Quek2007,Mowbray2008,Kaasbjerg2008} and better functionals for the description of molecule-substrate interfaces,\cite{Baer2010} also strongly suggest that some of the problematic issues can be handled satisfactorily.

Moreover, the problems with accounting for dynamical Coulomb correlations are less pronounced in the strong-coupling regime and when studying off-resonant transport, which is why NEGF+DFT performs well there. But even outside these regimes, while \emph{e.g.}~exact peak positions may not be accurate, the qualitative trends obtained by the method are still quite useful for understanding transport through molecular devices.

It is in this spirit that we report on an efficient NEGF+DFT  method implemented in the BAND periodic-system DFT code,\cite{Velde1991,Wiesenekker1991,Velde2001} (sister-code to the ADF molecular DFT code) which allows us to study novel single-molecule systems. A number of unique features of BAND, in particular the freedom to choose the number of dimensions in which periodicity is imposed, enables us to perform accurate modeling of the contacts, as well as of the electric potential in the presence of a gate. This will help to resolve a number of the issues critical to the full understanding of experimental results, although a full treatment of the Coulomb potential in the presence of a gate electrode has not yet been implemented in our method. \\


In the following section (\ref{Formalism}), we will briefly review the NEGF formalism underlying our approach, which has been described in detail elsewhere.\cite{Datta1997} Section \ref{Implementation} then discusses our implementation schematically, commenting on some subtle points relating to the peculiarities of BAND and critical to the efficiency of the calculations. In particular, we discuss the partitioning of the model system in \ref{Partitioning}, the treatment of the Hamiltonian obtained from the periodic band-structure calculation in \ref{Tight-binding}, and how we obtain surface Green's functions and self-energies, coupling the infinite leads to an ``extended molecule,'' in \ref{Surface}. In section \ref{Alignment} (and appendix \ref{AlignmentTests}) we then discuss the technical issue of the alignment of the potential between computational stages.
The details of the non-equilibrium calculations are treated in section \ref{NEqPMat}.

Once the alignment has been determined self-consistently, we proceed to the transport-calculation proper, for a molecular system of interest.  Calculations we have performed to validate our code are discussed in sections \ref{Transport1D} and \ref{Transport3D}.  In the 1D case we present benchmark results on  Li and Al chain systems, and in the general 3D case we discuss our study of transport through benzenedithiol and oligo-phenylene-ethynylene-dithiols, as archetypal molecular wires, between Au electrodes, which form a family of structurally related junctions. This is reflected in the character of their zero-bias transmission near the Fermi level $\epsilon_f$, where there is a progressive closing of the transport gap mirroring the decreasing gas-phase HOMO-LUMO gap. We further examine the details of the symmetries of the orbitals which are found to couple in transport, finding them again related across the family, and finally we show the results of calculations under moderate bias, which illustrates the screening effect of the electrodes and supports the common-practice of using the zero-bias transmission $T(\epsilon)$ for systems under low-bias.

\section{Overview of the NEGF+DFT Approach}\label{Formalism}

\begin{figure}
\includegraphics[width=\columnwidth]{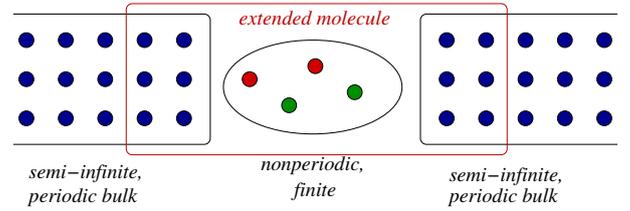}
\caption{Schematic geometry of the extended molecule (physical molecule and portion of the contacts to which it is attached) and semi-infinite bulk portion of the contacts.}\label{fg:extendedmolecule}
\end{figure}

In order to model the molecular system and the contacts to which it is connected using \emph{ab initio} techniques, we need to reduce the size of the system being modeled from a molecule between infinitely large contacts to something more manageable. To this end we use the ``extended molecule'' (EM) scheme,\cite{Xue2002,Brandbyge2002} illustrated in Fig.~\ref{fg:extendedmolecule}. The system is partitioned into a central extended molecule comprised of the actual molecule and some connecting parts of the leads on each side. This extended molecule, in turn, is then connected to true semi-infinite bulk leads via a well-defined metal-metal interface. Thus, in our model the leads are described by a finite-dimensional Hamiltonian $H$ corresponding to a portion of true bulk metal, corrected by a self-energy $\Sigma$ containing the response of the leads.

The key benefits of modeling the system in this way are first that we are able to place the interface between the leads and the active portion of the single-molecule system between metal layers, an interface which is much better understood than the complex molecule-metal binding geometries which may occur. The details of these binding geometries may then be varied, without the need to recalculate the contributions from the bulk contacts. 
A subtler point is that, as argued by Evers \emph{et al.}\cite{Rothig2006,Arnold2007}, this approach allows us to increase the size of the extended molecule in order to test convergence to transport properties which correspond to truly bulk-reservoir electrodes, which they have shown for tight-binding chains and cluster Au electrodes of varying sizes.
Furthermore, the metallic parts of the extended molecule allow us to take simple polarization effects in the leads into account. Finally, the approach allows us to derive an expression for the propagator for the entire extended molecule in a simple way, which we discuss in section \ref{Implementation}.

Our treatment of the metallic contacts is a Kohn-Sham based periodic band-structure calculation, as implemented in the BAND density-functional code.\cite{Velde1991,Velde2001} The code uses localized basis-functions, of either Slater-type orbital (STO) or numerical atomic orbital (NAO) type, usually complemented by frozen-core approximations of the inner electron shells of the atoms in the system. Smooth radial confinement of the basis functions can be applied using a Fermi-Dirac function of the distance from the nucleus. The code also supports variable periodicity, ranging from 0D (none), 1D (chain), 2D (slab) and finally to 3D (bulk) geometries.\\

First, a calculation is performed in the band-structure DFT code BAND to obtain the Fock matrix for a bulk unit cell, from the self-consistent density which enters the Kohn-Sham Hamiltonian $\mathcal{H}^\text{\tiny KS}$:
\begin{align}\label{ks} 
\mathcal{H}^\text{KS}(\bm{r}) = &-\frac{1}{2} \nabla^2 - \sum_n \frac{Z_n}{|\bm{r} - \bm{r}_n|} + V_\text{H}[n](\bm{r}) + V_\text{xc}[n](\bm{r})\;,
\end{align}
where $V_\text{H}[n](\bm{r})$ (the Hartree potential) is the solution of the Poisson equation with boundary conditions corresponding to the chemical potentials in the electrodes. The density is constructed from the occupied Kohn-Sham orbitals as $n(\bm{r}) = \sum_{i_\text{occ}} \left| \phi_i(\bm{r}) \right|^2$.

In the following, it should be kept in mind that the Kohn-Sham Hamiltonian depends on the bias voltage and the electron density, though these dependencies are usually omitted for compactness of notation. In BAND we represent the Hamiltonian \eqref{ks} in a non-orthogonal basis, and so find the Fock matrix as $H^\text{\tiny KS} = [ \langle\phi_i|\mathcal{H}^\text{\tiny KS}|\phi_j\rangle]$  and the overlap matrix as $S^\text{\tiny KS} = [ \langle\phi_i|\phi_j\rangle]$. From the Fock and overlap matrices for the bulk leads we calculate the surface Green's functions $g_c(\epsilon)$ by an efficient recursion algorithm.\cite{Sancho1985,Henk1993} We then find the corresponding self-energies $\Sigma_{1,2}(\epsilon)\sim\tau g_c(\epsilon) \tau^\dagger$ of the contacts (1,2 for source and drain, respectively and $\tau$ the coupling between layers in the leads), which are then combined with the Hamiltonian of the extended molecule to find the full Green's function:
\begin{equation}\label{GF1}
G_\text{\tiny EM}(\epsilon) = \biggl( \epsilon\,S_\text{\tiny EM} - H_\text{\tiny EM} -\left(\Sigma_1(\epsilon)+\Sigma_2(\epsilon)\right)\biggr)^{-1}\;. 
\end{equation} 
where $G(\epsilon),\;\Sigma_{1,2}(\epsilon)$ refer to the retarded (causal) Green's function and self-energies, respectively.
The Green's functions are then used in a modified self-consistent field (SCF) approach based on the density derived from the Green's function:  
\begin{align}\label{SCF2}
\nonumber \phi^{(0)}_j(\bm{r}) \rightarrow n^{(0)}(\bm{r}) \rightarrow \;&H^\text{\tiny KS} +\Sigma_1 + \Sigma_2 \rightarrow G(\epsilon) \rightarrow \rho \rightarrow n^{(1)}(\bm{r}) \rightarrow \\ \nonumber &H^\text{\tiny KS} +\Sigma_1 + \Sigma_2 \rightarrow G(\epsilon) \rightarrow \rho \rightarrow n^{(2)}(\bm{r}) \rightarrow\\ &\ldots
\end{align}
which may be compared to the usual SCF cycle in DFT:
\begin{align*}
 \phi^{(0)}_j(\bm{r}) \rightarrow n^{(0)}(\bm{r}) \rightarrow H^\text{\tiny KS} \rightarrow \phi^{(1)}_j(\bm{r})
\rightarrow n^{(1)}(\bm{r}) \rightarrow H^\text{\tiny KS} \ldots
\end{align*}
If we assume that the contacts each couple only with the central extended molecule, then we can simplify the general Green's function formalism and obtain the density matrix from the Green's function at each iteration as an integral over the real-valued energies:
\begin{align}\label{density2}
\nonumber \rho = \frac{1}{2 \pi} \int \mathrm{d}\epsilon\; [\; &G(\epsilon) \Gamma_1(\epsilon) G^{\dagger}(\epsilon) f(\epsilon,\mu_1) \\+ &G(\epsilon) \Gamma_2(\epsilon) G^{\dagger}(\epsilon) f(\epsilon,\mu_2)\;]\;.
\end{align} 
$\Gamma(\epsilon)$ is defined as $i\left( \Sigma(\epsilon)-\Sigma^\dagger(\epsilon)\right)$ for each contact. $\mu_1$ and $\mu_2$ are the chemical potential of source and drain electrodes. The bias voltage follows as $V_b = \frac{\mu_1-\mu_2}{e}$. 

This is general to the case of differing chemical potentials (\emph{e.g.} biased devices or different contact materials), but doesn't take \emph{e.g.} direct coupling between leads into account. In the equilibrium case with a single chemical potential in both leads, the expression further simplifies to:
\begin{align}\label{density1}
\rho = -\frac{1}{\pi}\int \mathrm{d}\epsilon\; \im{G(\epsilon)} f(\epsilon,\mu)\;.
\end{align}

When the SCF cycle \eqref{SCF2} converges, some interesting properties of the molecular system may be evaluated, using the converged $G(\epsilon)$ to obtain \emph{e.g.} the density of states (DOS) by:
\begin{align}\label{DOS}
D(\epsilon) &= -\frac{1}{ \pi} \tr{ \im{ \,G(\epsilon)\,}\,S }\;,
\end{align}
and the current by a Landauer-like expression, with $T(\epsilon) \sim \trr{ \Gamma_1 G \Gamma_2 G^{\dagger} }$ such that:
\begin{align}\label{Current}
I &= \frac{2e}{h} \int \mathrm{d}\epsilon\; \trr{ \Gamma_1(\epsilon) G(\epsilon) \Gamma_2(\epsilon) G^{\dagger}(\epsilon) } \left( f(\epsilon,\mu_1) - f(\epsilon,\mu_2) \right)\;.
\end{align}
The integral above is over the real line, but can be performed much more efficiently by using analytic continuation and complex contour integration.\cite{Zeller1982}\\

An important technical issue is that the offset of the potential cannot be expected to be the same in the junction geometry as that used for the bulk metal calculation for the contacts, due both to the controlled approximation in the tight-binding fit, and the much larger intrinsic issue of a floating-potential effect in the periodic band-structure code. The latter arises because in any band-structure DFT code the potential, and thus the Hamiltonian, are only determined up to some additive constant, \emph{i.e.} $H$ and $H+\dphi\, S$ give the same spectrum for constant offset $\dphi$. However, as our approach to transport involves several stages (bulk calculation of contacts, self-energy calculation, self-consistent alignment and transport calculation), we must take care to ensure that the (arbitrary) offset in the potential is consistent across all stages.

To find the offset  $\dphi$, an alignment calculation is carried out next, such that for the contacts in the transport calculation, it holds that:
\begin{align*}
H^\text{\tiny KS} \approx H^\text{\tiny TB} + \dphi\cdot S^\text{\tiny TB}\;,
\end{align*}
to within some acceptable tolerance. Once converged, we proceed to the calculation of an arbitrary molecular system, which may be under bias.

\section{Overview of the Implementation}\label{Implementation}

\subsection{System Partitioning}\label{Partitioning}

\begin{figure}
\includegraphics[width=\columnwidth]{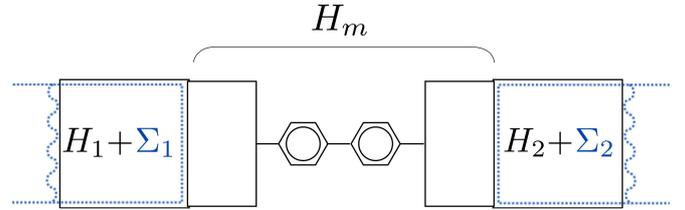}
\caption{Schematic geometry of the extended molecule and semi-infinite bulk portion of the contacts indicating the regions on which each operator in the formalism is active}\label{fg:extendedmolecule2}
\end{figure}

The partitioning illustrated in Fig.~\ref{fg:extendedmolecule} is made precise in terms of the operators in the formalism in Fig.~\ref{fg:extendedmolecule2}. The partitioning of the Hamiltonian into extended molecule ($H_m$) plus contacts ($H_c = H_1\oplus H_2$) is as follows:
\begin{align}\label{partitioning}
H &= \left(
      \begin{array}{ccc}
        H_m & \tau_1^\dag & \tau_2^\dag \\
        \tau_1 & H_1 & 0 \\
        \tau_2 & 0 & H_2 \\
      \end{array}
    \right) \equiv
    \left(
      \begin{array}{cc}
        H_m & \tau^\dag \\
        \tau & H_c \\
      \end{array}
    \right) 
\end{align} 
such that the Green's function for the entire system is determined by inverting:
\begin{align}\label{matGF}
(\epsilon\,S - H)G(\epsilon)=I\;.
\end{align}

The specific approximation behind this partitioning is that we require the full Hamiltonian for contacts and extended molecule to be correct in the Kohn-Sham sense at each iteration, while the density matrix need only be correct for the extended molecule. This is justified by its spatial separation from the (expected) field- and charge-errors at the edges of the finite system. The errors there, in turn, are minimized by replacing $H_{1,2}$ by the stored bulk operators, as we discuss in section \ref{Alignment}.

\subsection{Tight-binding Representation of $H^\text{\tiny KS}$}\label{Tight-binding}

From the bulk calculation, the Fock matrix is known with respect to BAND's basis functions for many Bloch boundary conditions, represented by a dense set of Bloch wavevectors $\bm{k}$ inside the Brillouin zone. We search for a tight-binding Fock matrix, given in terms of BAND's localized  
basis functions, which reproduces BAND's Fock matrices as closely as possible. 

The basis functions
are denoted $\left| \phi_i^\alpha\rangle\right.$, where $i$ denotes a unit cell and $\alpha$ denotes a particular orbital type:
\[
\langle \bm{r} | \phi^\alpha_i \rangle = \phi^\alpha (\bm{r} - \bm{r}^\alpha_i). 
\]
Here, $\bm{r}^\alpha_i$ is the position of the atom in unit cell $i$ about which the $\alpha$-orbital is centered. Taking the cell index $i=0$ without loss of generality, we denote the real-space tight-binding matrix elements in relation to $k$-space as:
\begin{align}
H^\text{\tiny TB}_{\alpha\beta}(\bm{k}) &= \sum_{\bm{R}_j} \; H^\text{\tiny TB}_{i\alpha,j\beta} \; e^{i\bm{k}\cdot\bm{R}_j}\;,
\end{align}
The requirement that the matrix on the left-hand side is equivalent to BAND's Fock operator suggests that:
\begin{align*}
F = \sum_{\alpha\beta} \sum_l \left|H_{\alpha\beta}^\text{\tiny TB}(\bm{k}_l) - H_{\alpha\beta}^\text{\tiny KS}(\bm{k}_l)\right|^2\;,
\end{align*} 
be minimized with respect to the real-space matrix elements $H^\text{TB}_{i\alpha,j\beta}$.

This minimization is done using the L-BFGS algorithm,\cite{Press2007}  with a cut-off radius
which is increased as needed to meet a user-specified tolerance for the fit.
As each element of the tight-binding fit is calculated using only the lattice positions and the reference value of the Fock matrix, it can be calculated independently, so that this calculation is trivially parallelized, with linear scaling in the number of cores.

We emphasize, however, that we make a tight-binding \emph{fit} $H^\text{\tiny{TB}}$ of $H^\text{\tiny KS}$ in terms of the (known) lattice of the contact, as opposed to switching to a traditional tight-binding \emph{model} for the electronic structure. These TB Fock and overlap matrices, together with the lattice, then form the input for our calculation of the surface Green's functions. \\

\begin{figure}
 \includegraphics[width=.8\columnwidth]{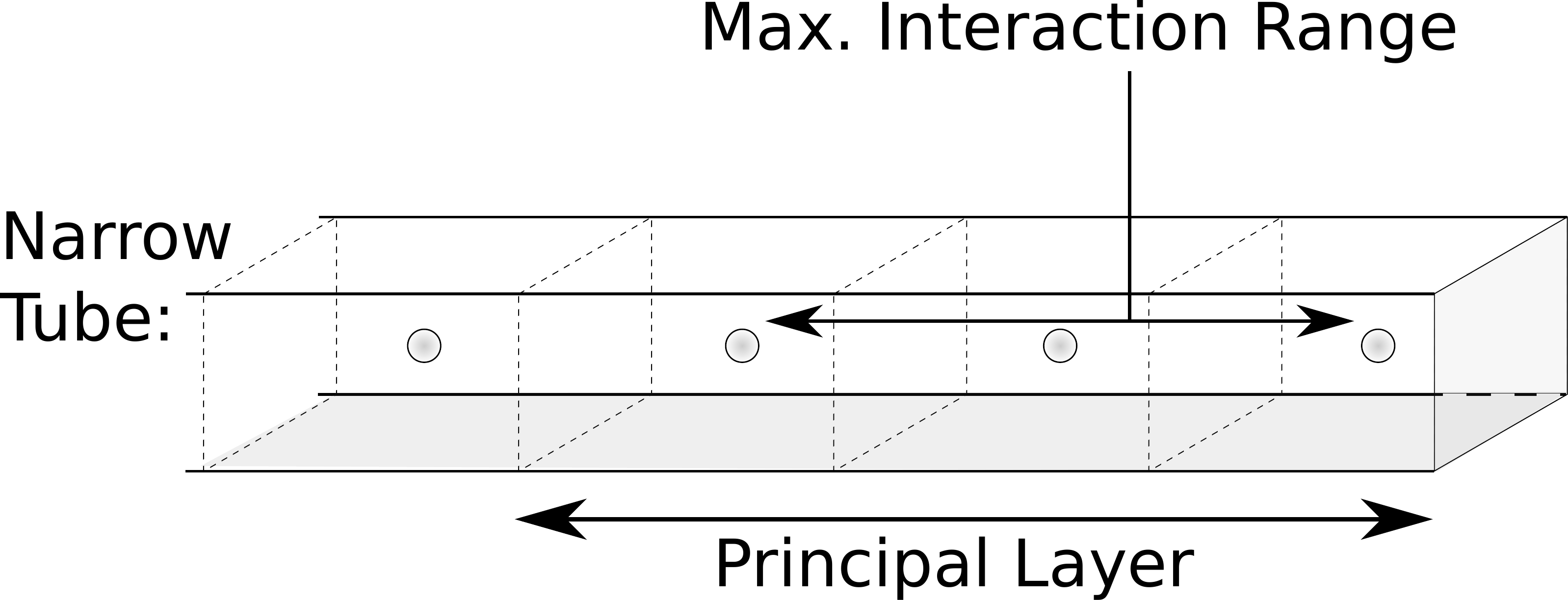}
\caption{Narrow tube (NT) in the infinite contact based on a single unit cell. Partitioning based on principal layers, which due to electronic screening interact only with neighboring principal layers (discussed in section \ref{Surface}).}\label{fg:players}
\end{figure}

\begin{figure}
 \includegraphics[width=.8\columnwidth]{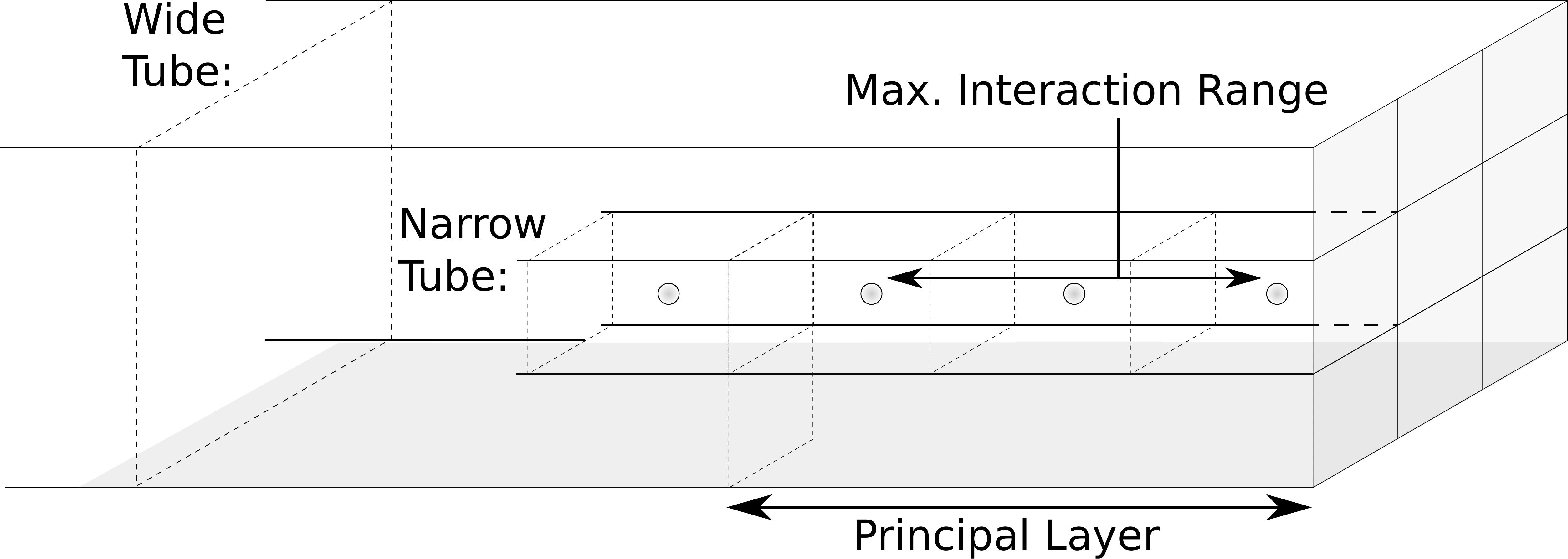}
\caption{Wide tube (WT) in the infinite contact based on an $\ell\times\ell$ grid of unit cells in the surface plane. Compare Fig.~\ref{fg:players} for the relation with the narrow tube. }\label{fg:tube}
\end{figure}

We now have a set of Fock and overlap matrices corresponding to $\bm{k}$-vectors in the Brillouin zone, from which we can calculate self-energies, for a `tube' extending into the contact as in Fig.~\ref{fg:players}. From these, we can calculate self-energies on the same grid of $\bm{k}$-vectors. We assume the number of $\bm{k}$ vectors in the Brillouin zone to be $N^2$.
From these matrices, it is then possible to calculate their counterparts for a wider tube with periodic boundary conditions, see Fig.~\ref{fg:tube}. We now denote the original (narrow) tube by NT, and the wide tube as WT.

We assume that the 2D unit cell of the wide tube is an integer multiple ($\ell \times \ell$) of that of the narrow tube.
Within the narrow tube, we can write the Fourier transform of the matrix $M$ (which may be the Hamiltonian or the overlap matrix) for any wave vector $\bm{k}$ as  
\[
M^\text{NT}_{\alpha \beta} (\bm{k}) = \sum_{\bm{r}_\|} M_{\alpha,\beta} (\bm{r}_\|) e^{i \bm{k} \cdot \bm{r}_\|},
\]
where the sum $\sum_{\bm{r}_\|}$ is over \emph{all} relative positions $\bm{r}_i^\alpha-\bm{r}_j^\beta$ within the large volume 
corresponding to the $N^2$ $\bm{k}$-vectors in the two-dimensional BZ in the surface plane. 

Taking the wide tube with periodic boundary conditions (i.e.\ at the $\Gamma$-point), we have
\[
M^\text{WT}_{\alpha\beta} (\bm{r}_\|) = \sum_{\bm{R}^\text{WT}} M_{\alpha \beta} (\bm{r}_\| + \bm{R}^\text{WT})\;,
\]
where $\bm{R}^\text{WT}$ is any $m\cdot\bm{a}_1^w+n\cdot\bm{a}_2^w$,  linear combination of wide-tube basis vectors ($\bm{a}_i^w = \ell\bm{a}_i$ in terms of the NT basis vectors).
This equation expresses the periodicity of $M^\text{WT}$.
On the other hand we can write:
\[
M_{\alpha\beta}(\bm{r}_\| + \bm{R}^\text{WT}) = \frac{(2\pi)^2}{\Omega} \int d^2 k\; M_{\alpha\beta} (\bm{k}) 
e^{i \bm{k} \cdot (\bm{r}_\|+\bm{R}^\text{WT})}.
\]

Combining these two ingredients, we can write
\[
M^\text{WT}_{\alpha\beta} (\bm{r}) = \sum_{\bm{R}^\text{WT}} \frac{1}{N^2} \sum_{\bm{k}_\text{BZ}} M_{\alpha\beta}(\bm{k}_\text{BZ}) 
e^{i \bm{k} \cdot \left(\bm{r}_\| + \bm{R}^\text{WT} \right)}.
\]
The sum over the wide tube vectors $\bm{R}^\text{WT}$ singles out the reciprocal lattice vectors (lying inside the NT Brillouin zone)
of the wide tube:
\[
\sum_{\bm{R}^\text{WT}} e^{i \bm{k} \cdot \bm{R}^\text{WT}} = \frac{N^2}{\ell^2} \sum_{\bm{K}^\text{WT} \,\in\, BZ^\text{NT}} 
\delta(\bm{k} - \bm{K}^\text{WT}),
\]
from which we immediately have:
\[
M^\text{WT}_{\alpha\beta} (\bm{r}_\|) = \frac{1}{\ell^2} \sum_{\bm{K}^\text{WT} \,\in\, \text{BZ}^\text{NT}} M^\text{NT}_{\alpha\beta}(\bm{K}^\text{WT}) 
e^{i \bm{K}^\text{WT} \cdot \bm{r}_\|}\;.
\]
This expression tells us how to obtain the matrix elements between any two points inside the wide tube from the matrix elements on
the reciprocal lattice points of the WT, lying inside the Brillouin zone of the NT.

\subsection{Surface Green's Function and Self-Energies}\label{Surface}

Equation \eqref{matGF} for the matrix Green's function in the case of a non-orthogonal basis 
yields the matrix relation:
\begin{align}
\left( \begin{array}{cc}
I_m & 0  \\
0 & I_c  \\ \end{array} \right) &=
\left( \begin{array}{cc}
\epsilon S_m - H_m & \epsilon S_{\tau}^{\dagger} - \tau^{\dagger}  \\
\epsilon S_{\tau} - \tau & \epsilon S_c - H_c  \\ \end{array} \right)
\left( \begin{array}{cc}
G_m & G_{mc}  \\
G_{cm} & G_c  \\ \end{array} \right)
\end{align}
(omitting the $\epsilon$-dependence from the Green's functions on the right-hand side) from which follow closed expressions for the propagator $G_m$ of the extended molecule using self-energies, in the presence of a contact:
\begin{align}
G_m(\epsilon) &= \left( \epsilon S_m - H_m - (\epsilon S_{\tau}^{\dagger} - \tau^{\dagger}) G_c(\epsilon) (\epsilon S_{\tau} - \tau) \right) ^{-1}\;, 
\end{align}
and we identify:
\begin{align*}
\Sigma_c(\epsilon) \equiv (\epsilon S_{\tau}^{\dagger} - \tau^{\dagger}) G_c(\epsilon) (\epsilon S_{\tau} - \tau)
\end{align*}
as the self-energy of the contacts, which may be split as $\Sigma_1+\Sigma_2$.\footnote{In the literature $\Sigma = \tau g \tau^\dagger$ is often used as the definition of the self-energy; then $\tau$ corresponds precisely to our $(\epsilon S_{\tau} - \tau)$ expression for the coupling of device and contact.}  For two contacts we specify the propagator further as $G(\epsilon)=\left( \epsilon S_m - H_m - \Sigma_1 - \Sigma_2 \right)^{-1}$, and it is this subsystem which we subsequently focus on.\\ 

Two further remarks are important before continuing. First, we note that the determination of $\Sigma_c$ only requires knowing the surface couplings in $(G_{\text{c}})_{i j\, \in\,\text{surf}}$, which makes this practical to implement in a DFT code with localized basis functions (such as BAND). Second, we note that while the above is an exact description within the limits of the one-electron picture (\emph{i.e.} neglecting electron-electron interaction beyond the mean-field level), in practice when calculating by an approximate method such as an actual DFT implementation, we need to be aware of the consequences of the limited spatial extent of the extended molecule, which may be felt by the central region due to the Hartree term in the potential if it is insufficiently screened. Generally, for metals, the screening is strong enough to justify the approach for contacts of a few atomic layers.\\

Now, our approach to obtaining the Green's function $G$ relies on the fact that the metallic system has a finite interaction-range in real-space due to electronic screening.  This implies a local-- and neighbor-coupling structure of the Fock matrices which is tridiagonally structured as $( \tau,\; h,\; \tau^\dagger )$ in a basis organized into adjacent layers of atoms, and similarly tridiagonally structured as $( s_\tau,\; s,\; s_\tau^\dagger )$ for the overlap matrix.

We introduce the concept of ``principal layers,'' because it is well known that electronic screening limits the interaction range of the Coulomb potential to just a few atomic layers in a metal.\cite{Ashcroft1976} This implies that we can give a description in terms of blocks of 3-4 atomic layers called a ``principal layer,'' which interact only with the neighboring principal layers. Together with the use of localized basis functions, this allows us to use the structure of a Hamiltonian matrix as in equation \eqref{H_principal_layers} corresponding to principal and adjacent (interacting) layers, as illustrated in Fig.~\ref{fg:players}. 
\begin{align}\label{H_principal_layers}
H=\left(
  \begin{array}{ccccc}
    h & \tau^\dagger & 0 & \cdots &\\
    \tau & h & \tau^\dagger & 0  & \cdots\\
    0 & \tau & h & \tau^\dagger & 0\\
    \vdots & 0 & \ddots & \ddots & \ddots  \\
  \end{array}
\right)\;.
\end{align}

We consider the metal as being composed of an infinite number of layers in space, and then find the relation between elements of the Green's function for $2^k$ and $2^{k+1}$ principal layers by recursively eliminating the layers in between.\cite{Sancho1985,Henk1993}

\begin{figure}
\includegraphics[width=\columnwidth]{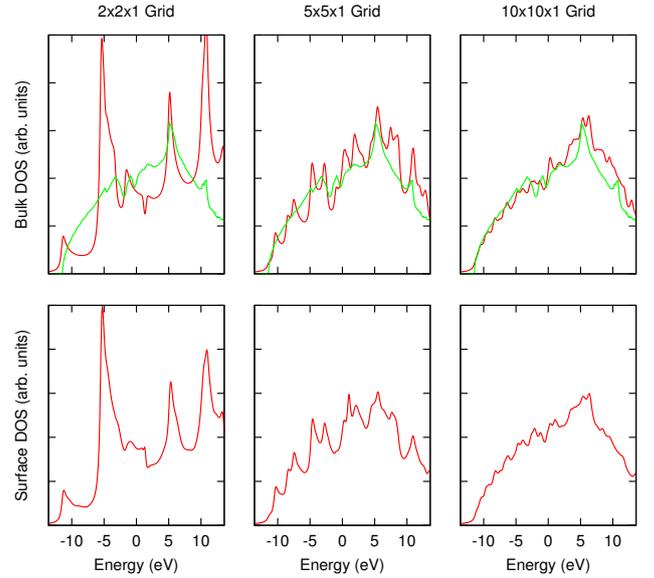}
\caption{Convergence of the bulk aluminum DOS with increasingly dense grid in $k$-space.
Top: bulk DOS from the Green's function ({\bf red}) compared with normal DFT calculations ({\bf green}): convergence for finer $k$-grids. Bottom: surface DOS for a [111]--cut surface (from the surface Green's function), which cannot be obtained directly from DFT. DFT calculations performed with the LDA functional, using a DZP-quality basis-set.}\label{fg:dosconverges}
\end{figure}

From this we obtain the Green's function for the surface and the bulk of an infinite contact, and can study the latter's convergence with respect to the bulk calculation in BAND. This method is easily extended to evaluation over a Monkhorst-Pack grid\cite{Monkhorst1976} in $k$-space, and parallelized in energy.
This approach converges quickly, and a sample calculation of the surface and bulk DOS is illustrated in Fig.~\ref{fg:dosconverges} for different grid densities in $k$-space in the plane of the contacts. In principle this $k$-space dependence also carries over to the alignment and transport calculations, and in the previous section we discussed a method for the construction of an expanded self-energy for the contacts. However, we will present only calculations in the $\Gamma$-point approximation in the remainder of this paper.

The key computational steps in this stage of the calculation are the complex contour integrals over the Green's functions (expressions \eqref{density2}--\eqref{density1}), which are a much more efficient way\cite{Zeller1982} to evaluate the density matrix from the Green's function than direct integration over the real line. The reason for the latter is that in general $G(\epsilon)$ may have poles quite near the real axis, necessitating a very dense integration grid, while the contour may be taken safely away from these in the upper half of the complex plane, drastically reducing the computational effort. 

We pre-calculate the contours themselves and the corresponding self-energies $\Sigma_{1,2}(\epsilon_i)$ over all points on the contour $\{\epsilon_i\}$, given that the self-energies can be calculated independently for each energy point on the contour.

\subsection{Potential Alignment and Determination of the Fermi Level}\label{Alignment}

As noted above, the zero of the potential is not uniquely determined in this type of DFT calculation. As our approach to transport  involves a sequence of relatively independent computations (bulk calculation of contacts, self-energy calculation, self-consistent alignment and transport calculation), we must ensure that the (arbitrary) offset in the potential is consistent across all stages, keeping in mind that the self-energies also implicitly reference the chemical potential of their respective contact: $\Sigma_{1,2}(\epsilon_i;\mu_{1,2})$.
A number of codes take different approaches to this,\cite{Stokbro2003a,Rocha2006} but we are not aware of any approach that has handled the problem self-consistently to date.

In order to ensure the alignment of the potentials in the leads (and their self-energies) with those of the extended molecule, we first note that there is a natural criterion for determining the offset: the charge neutrality of bulk material. Clearly, the chemical potential is directly related to the number of electrons in the metal. Consequently, when the unbiased extended molecule is itself composed of the same material as the contacts, we can self-consistently determine the offset by requiring the (valence) charge on the extended molecule to equal the bulk (valence) charge for the same number of atoms. We tune this charge by iteratively shifting the potential during the SCF until the criterion is met. We have also implemented a novel constrained-DIIS (CDIIS) scheme in our code to accelerate the convergence of this alignment procedure for difficult systems; this is briefly outlined in appendix \ref{DIIS}.\\

\begin{figure}
 \includegraphics[width=.98\columnwidth]{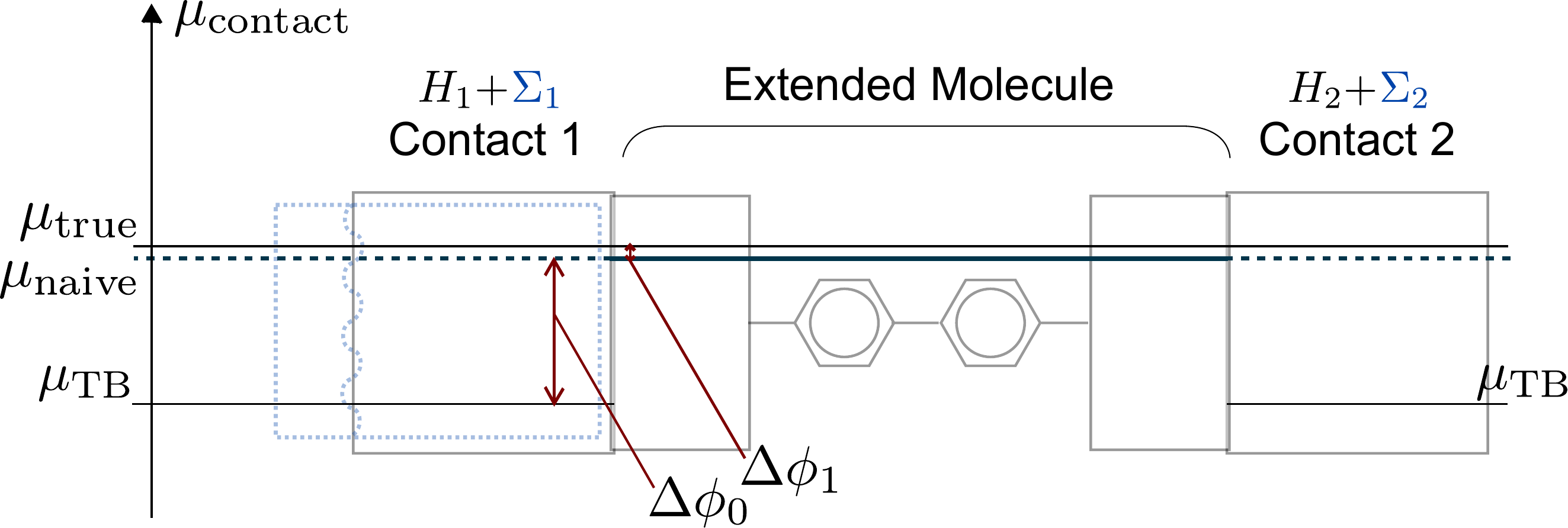}
\caption{Differing computational stages' zeros of the potential induce differing chemical potentials which are reconciled in alignment. Shown is the analog of Fig.~\ref{fg:extendedmolecule2}, where we now include the contact offset $\dphi_0$ and fine-tuning of charge-neutrality $\dphi_1$ as discussed, yielding a consistent $\mu_\text{true}$ after alignment.}\label{fg:potentialzeros}
\end{figure}

Our approach is to split the shift into the two components illustrated in Fig.~\ref{fg:potentialzeros}. The first is the offset between the bulk run (periodic cell, used to construct the self-energies for the semi-infinite contacts) and the alignment run (longitudinally aperiodic transport geometry composed of contacts + extended molecule). This offset is estimated at each iteration as follows:
\begin{align}
\dphi_0 = \frac{1}{n_\text{bas}}\displaystyle\sum_{i\in\text{C1}\oplus\text{C2}} \frac{\left(H_{ii}^\text{\tiny TB}-H_{ii}^\text{\tiny KS}\right) }{ S_{ii} }\;,
\end{align}
where $n_\text{bas} = n_\text{bas}^\text{C1} + n_\text{bas}^\text{C2}$ is the dimension of the basis of the Hamiltonian for the two contacts and $H^\text{\tiny TB}$ and $H^\text{\tiny KS}$ refer to the tight-binding representation of these bulk contact Hamiltonians and the transport geometry's contact Hamiltonians respectively (see Fig.~\ref{fg:extendedmolecule}).
We now shift the system by the offset: $H^\text{\tiny KS} \rightarrow H^\text{\tiny KS} + \dphi_0\, S^\text{\tiny KS}$. Next, we overwrite the Hamiltonian and overlap matrix $S^\text{\tiny KS}$ of the contact regions by those obtained from the bulk calculations, which do not suffer from edge effects and are now aligned with the rest of the system. This is updated at each iteration, allowing for fluctuations in the density-dependent potential $V[n({\bf r})]({\bf r})$, such that the extended system is always as close as possible to be precisely aligned with the bulk contacts' potential.\footnote{Convergence of these shifts is a natural heuristic for the proper alignment of the system with this algorithm.}

The second shift is the correction to the extended system which brings the Fermi level into alignment with the implicit chemical potential encoded in the open-boundary self-energies. We obtain it by determining the density matrix from the Green's function for the extended system via equation \eqref{density1}, which yields the valence charge over the extended molecule region by tracing over the relevant basis functions as $\trr{\rho S}_\text{\tiny EM}$. This is in practice typically not charge-neutral. To achieve $Q_\text{\tiny EM} \rightarrow Q_\text{valence}$ we use an offset to ensure charge neutrality, which is calculated iteratively:
\begin{align}
\dphi_1\itr{k+1} = \dphi_1\itr{k} + \alpha\left(\tr{\rho S}_\text{\tiny EM} -Q_\text{valence}\right)
\end{align}
until convergence is achieved. The shifts $\dphi_1$ are applied to the entire system. Note that the parameter $\alpha$ has the dimensions of a capacitance, and indeed can be chosen proportional to the inverse of the density of states, evaluated at the Fermi-energy. In practice, such an implementation easily becomes numerically unstable, and we have opted for a fixed, small mixing parameter $\alpha$ instead.

Both shifts, $\dphi_0,\,\dphi_1$, must converge for the alignment stage to be considered successful. This procedure may be accelerated by extending the DIIS scheme\cite{Pulay1980} as outlined in appendix \ref{DIIS}, and we illustrate a representative performance of the accelerated method in Fig.~\ref{fg:Al.ShiftConvergence-cdiis} for a 1D Al chain, as well as in figures Figs.~\ref{fg:Au.ShiftConvergence1}-\ref{fg:Au.ShiftConvergence2} for Au contacts. Moreover, for each set of contacts we discuss, we check that re-running the alignment ``bulk'' geometry as a zero-bias transport calculation using the shifts calculated as static inputs indeed results in charge-neutrality in the extended molecule.\\

To summarize, when the SCF calculation converges, we have obtained two potential shifts: the contact shift $\dphi_0$ 
and the charge-neutrality shift $\dphi_1$. The former is a runtime iterative adjustment to ensure that the active region of the transporting system is aligned to the bulk Hamiltonians with which the contacts are overwritten, while the latter is a runtime constant which ensures that the potential of the entire system is such that a bulk extended molecule is precisely charge neutral. This in turn determines the Fermi level completely. The alignment calculation is separate from our transport calculations, and performed once for every new set of contacts.

The subtlety of our approach lies in realizing that by aligning the transport system to the bulk calculation, we tie it to the picture of contacts as reservoirs with well-defined chemical potentials. Consequently, from this point onward the Fermi level is no longer an estimated quantity, but an exactly known and fixed quantity, stemming directly from the bulk periodic contact calculation.\\

The procedure outlined here performs well in practice, correcting the offsets illustrated in Fig.~\ref{fg:potentialzeros}, and produces a PDOS on the extended molecule which matches the PDOS of the bulk contacts very well, see section \ref{Transport1D}. 
The resulting electronic structure, moreover, compares well with a bulk calculation of the true periodic system, as shown in appendix \ref{AlignmentValidation}.

\subsection{Non-equilibrium Calculations}\label{NEqPMat}

The procedure described in the previous section yields the ``total shift'' $\dphi=\dphi_0+\dphi_1$, and from this point on $\dphi_1$ is a constant shift applied at every iteration during a transport run. However, in order to treat the non-equilibrium transport case, we also need to consider both the effects of the applied bias and fields, and the calculation of the non-equilibrium density from the NEGF formalism.

The system in the transport run is, as we have outlined, shifted from the unmodified $H^\text{\tiny KS}$ for the extended molecule with contacts to the correct potential zero as $H+(\dphi_0+\dphi_1)S$. To this we add the bias $\phi_b(\bf r)$ and (possibly) gate $\phi_g(\bf r)$ fields being applied to extended molecule region.\footnote{Not to the deep contacts and self-energies, which are already at the right chemical potentials.} The potential profile implementing these fields is usually a ramp whose end points lie sufficiently far from the electrodes' surface for the layers of contact material in between to sufficiently screen the local distortions and produce the correct self-consistent potential drop within the extended-molecule region.\footnote{In our implementation a number of models in these two classes are available.} We discuss this further for the case of biased gold-phenyl-gold junctions in section \ref{bias} below.

In order to calculate the current in the presence of a bias voltage, we need to calculate the non-equilibrium density matrix. Our approach is splitting expression \eqref{density2} into the equilibrium term we have used thus far, and a new non-equilibrium correction, given by the following expressions \eqref{density3a}-\eqref{density3b}, analogous to the approach of Stokbro \emph{et al.}:\cite{Stokbro2003a}
\[
\rho = \frac{1}{2 \pi} \int \mathrm{d}\epsilon\; [G(\epsilon) \Gamma_1 G^{\dagger}(\epsilon) f(\epsilon,\mu_1) + G(\epsilon) \Gamma_2 G^{\dagger}(\epsilon) f(\epsilon,\mu_2)],
\]
which may be worked out to yield:
\begin{align}\label{density3a}
\nonumber \rho &= \underbrace{-\frac{1}{\pi}\int \mathrm{d}\epsilon\; f(\epsilon,\mu_1)\im{G(\epsilon)}}_{\rho^\text{A}_\text{eq}} \\
&+ \underbrace{\frac{1}{2\pi}\int \mathrm{d}\epsilon\; [f(\epsilon,\mu_1)-f(\epsilon,\mu_2)]G(\epsilon)\,\Gamma_2(\epsilon) \,G^\dagger(\epsilon)}_{\rho_\text{neq}^\text{A}}
\end{align}
or equivalently:
\begin{align}\label{density3b}
\nonumber \rho &=\underbrace{-\frac{1}{\pi}\int \mathrm{d}\epsilon\; f(\epsilon,\mu_2)\im{G(\epsilon)}}_{\rho^\text{B}_\text{eq}} \\
&+ \underbrace{\frac{1}{2\pi}\int \mathrm{d}\epsilon\; [f(\epsilon,\mu_2)-f(\epsilon,\mu_1)]G(\epsilon)\,\Gamma_1(\epsilon) \,G^\dagger(\epsilon)}_{\rho_\text{neq}^\text{B}}\;.
\end{align} 
$\rho_\text{eq,neq}^\text{A}$ and $\rho_\text{eq,neq}^\text{B}$ are two equivalent ways of obtaining the equilibrium and non-equilibrium density matrix, from which we obtain the terms by a weighted average.

The equilibrium terms' integrals may be evaluated as before by complex contour integration, but it is important to observe that the argument underpinning the analytic continuation into the complex plane was the localization of the poles of $G(\epsilon)$ in the lower half-plane. This is no longer true for the more complicated pole structure of terms like $G(\epsilon)\,\Gamma_1 \,G(\epsilon)^\dagger$, and so the non-equilibrium integral must be evaluated along a dense grid as near to the real axis as is reasonable, while avoiding numerical inaccuracies due to nearby poles.

\section{Transport in 1D Systems}\label{Transport1D}

\subsection{Li Chains}

\begin{figure}
 \includegraphics[width=.8\columnwidth]{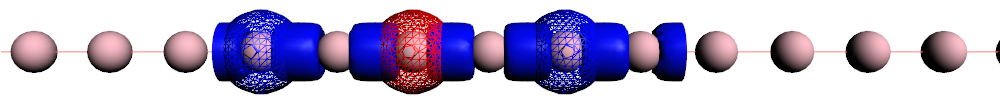}\vspace{.6cm}
 \includegraphics[width=\columnwidth]{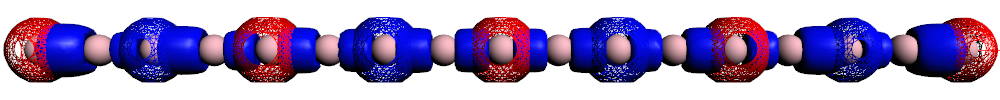}\vspace{.6cm}
\caption{Bulk charge density profile (solid) and HOMO wavefunction (hatched) calculated in a bulk 1D chain (top) and the converged alignment configuration for our finite extended system (bottom). Both were calculated using LDA and a SZ basis-set. Edge effects on the outer atoms on the finite chain are clearly visible, but in the inner extended-molecule region there is excellent agreement, including the 4-atom periodicity of the wavefunction.\label{fg:Li-Charge-Density}}
\end{figure}

In section \ref{Alignment} we outlined the use of a self-consistent procedure to fix the potential zero and the Fermi level consistently for the full system of extended molecule with contacts, by requiring charge neutrality of a bulk ``extended molecule.'' If we apply this to a 1D Li chain, we obtain a charge-density profile and HOMO wavefunction as illustrated in Fig.~\ref{fg:Li-Charge-Density}, where these are compared with the bulk result obtained from a conventional periodic DFT calculation, using the LDA exchange-correlation functional and a single-$\zeta$ (SZ) basis-set for Li. The lattice spacing (2.876 \AA) for the cell was obtained by energy minimization, using the same LDA functional and basis for consistency.

We point out that the exact 4-atom periodicity observed in Fig.~\ref{fg:Li-Charge-Density} is a signature result: we can derive the ``HOMO'' level (the highest occupied state in the infinite system) from a model of fermion sites in a finite chain by filling the outer $s$-orbital of Li at each site. Take a $M+1$-site chain of length $L$, with lattice constant $a$, which will hold $2(M+1)$ electrons. Each  electron will sit in a band formed by a standing wave pattern because of the periodic boundary conditions, and so:
\begin{align*}
\psi_n(x) = e^{i\, \pi n x/L}\;.
\end{align*}
The states may be labeled by the $2(M+1)$ set of $k_n = \pm\frac{\pi n}{L}$ instead, where the maximum value of $N$ is $M/2$ such that:
\begin{align}\label{lambda1}
\lambda_\text{max} &= \frac{2\pi}{k_\text{max}} \sim 4a
\end{align}
Consequently, the wavelength of the highest occupied mode in the infinite chain is 4 lattice spacings, which is exactly what we see in Fig.~\ref{fg:Li-Charge-Density}.

\begin{figure}
 \includegraphics[width=\columnwidth]{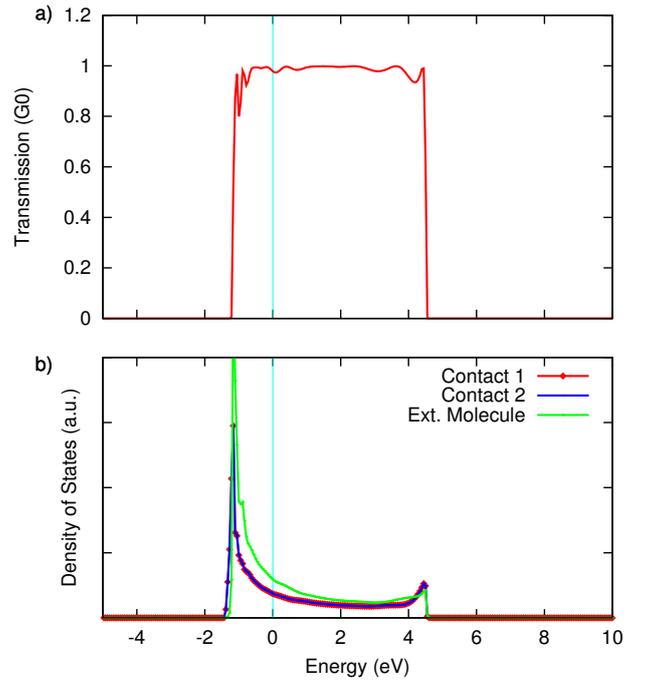}
\caption{(a) Zero-bias transmission through and (b) density of states of the 1D Li chain (using a SZ basis set), showing a single $s-$band providing transport as we expect, with a range in energy corresponding to the (projected) DOS on each part of the model structure.}
\label{fg:transmission-Li}
\end{figure}

The PDOS and transmission are illustrated in Fig.~\ref{fg:transmission-Li}, where we note that the transmission has a clear plateau at $G_0$ corresponding to the transmission through a single channel over the range of energy corresponding to nonzero density of states in the chain. The prediction of the wavefunction shape of the highest occupied mode is general for a single electron in an outermost atomic shell without degeneracy, such as the Li $2s$ orbital, as opposed to the Al chain (see below). We also note oscillations in the DOS and transmission, which likely reflect the finite extent of the contacts (the edges acting as scattering potentials), and the bad screening of a 1D chain in particular.\\

\begin{figure}
 \includegraphics[width=\columnwidth]{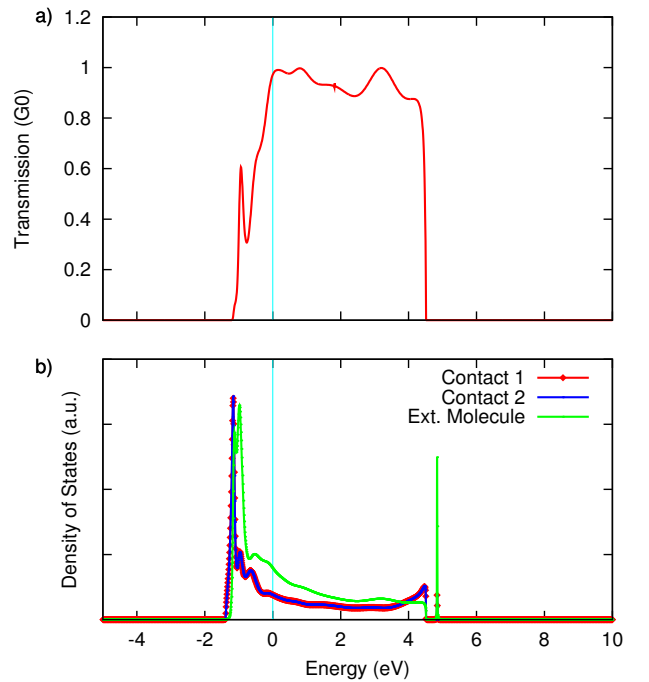}
\caption{(a) Zero-bias transmission through and (b) density of states of the 1D Li chain with an H$_2$ molecule (using a SZ basis set on Li, DZ basis on H$_2$). We see a decline of the transmission when a H$_2$ molecule is placed in the junction. Given the small size of the molecule, its primary effect is to slightly weaken the coupling in the chain.}
\label{fg:transmission-LiH2Li}
\end{figure}

We next performed a zero-bias transport calculation with a H$_2$ molecule placed in between the chain contacts. We calculate the transmission and PDOS for the structure under zero bias, see Fig.~\ref{fg:transmission-LiH2Li}.
We observe that the transmission is reduced in the presence of the H$_2$, an effect of H$_2$ partially interrupting the transport path through the s-orbitals in $1$s$^22$s$^1$ Li chain (transmission $\sim 1$, effectively an ideal Landauer conductance channel). The H$_2$ LUMO does have an s-orbital character, and so there is transport as it does in fact couple more broadly to a number of Li states, but its coupling to the Li is different, so that the transmission becomes somewhat attenuated at different energies along the band. 
On the high-energy side of the DOS plots, we also note a lone peak in the extended-molecule DOS: this corresponds to the LUMO orbital of H$_2$, and since its resonance is beyond the limited extent of the Li chain's DOS, it should not couple in transport. That is indeed what we observe in the transmission: the lack of a feature at the corresponding energy in the top panel of Fig.~\ref{fg:transmission-LiH2Li}. The HOMO and LUMO+1 levels, by contrast, are considerably further away, and do not couple at all.\\

A similar analysis can be performed for experimentally more realistic Al chains. The key difference is the addition of electrons, which now contribute both $s-$ and $p-$ bands for transport, as we discuss in appendix \ref{Al.Chains}, where we find otherwise very similar results.

\section{Transport in General 2-terminal Systems}\label{Transport3D}

\subsection{Building Au Contacts}

\begin{figure}
\subfloat[]{\includegraphics[width=.8\columnwidth]{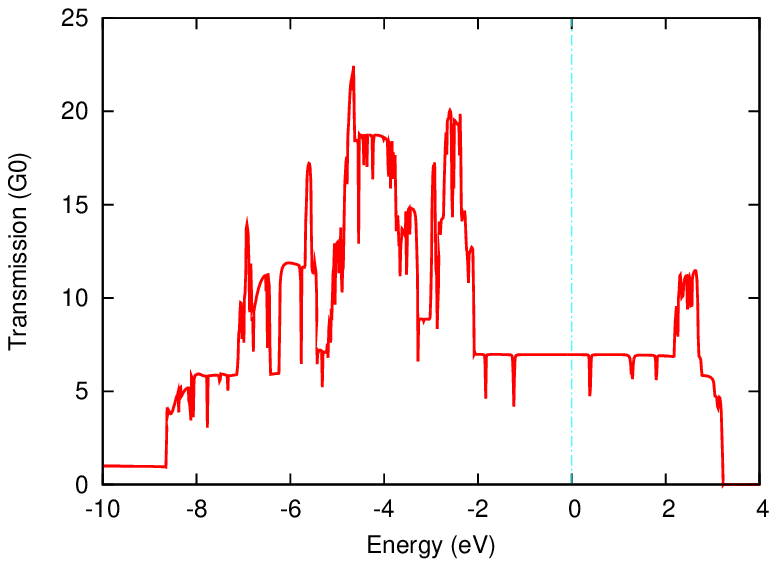}}\\
\subfloat[]{\includegraphics[width=.8\columnwidth]{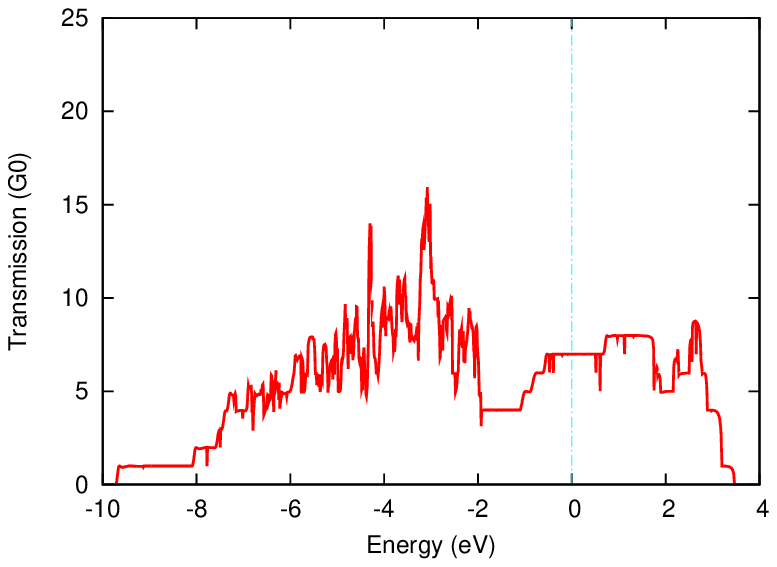}}
\caption{Transmission through the ``bulk'' Au contacts (LDA calculations using a SZ, 11$e$ basis per atom). {\bf a)} with and {\bf b)} without \pbc. Note the transmission plateaus at integer units of the conductance quantization $G_0$, as we expect from a system that is essentially a larger-diameter version of a single Au-atom chain. }\label{fg:Au.Transmissions}
\end{figure}

\begin{figure}
\subfloat[]{\includegraphics[width=.8\columnwidth]{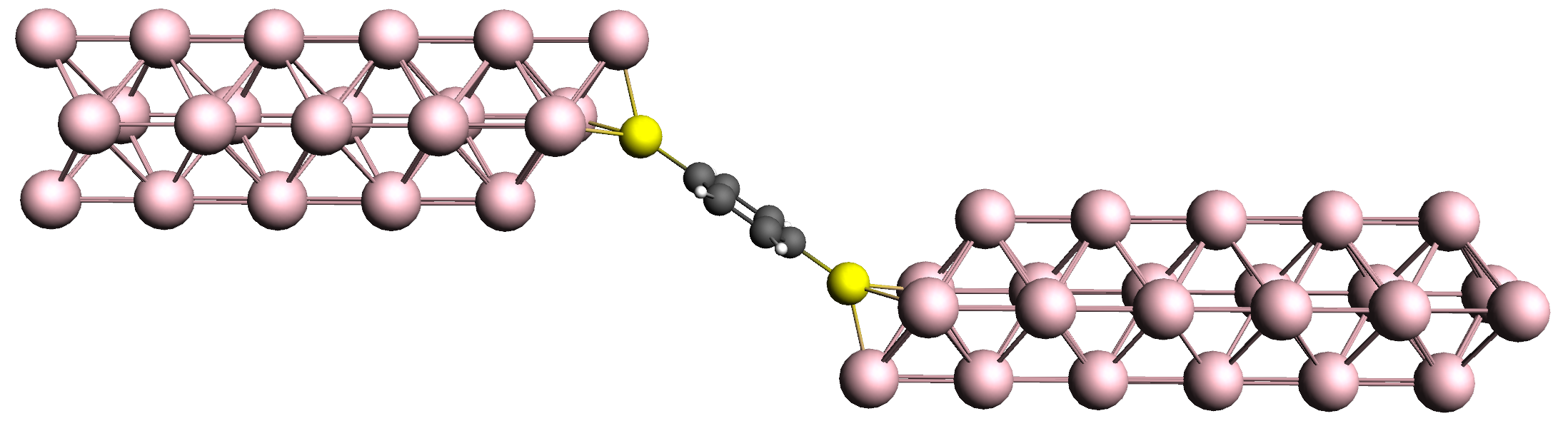}}\\
\subfloat[]{\includegraphics[width=.8\columnwidth]{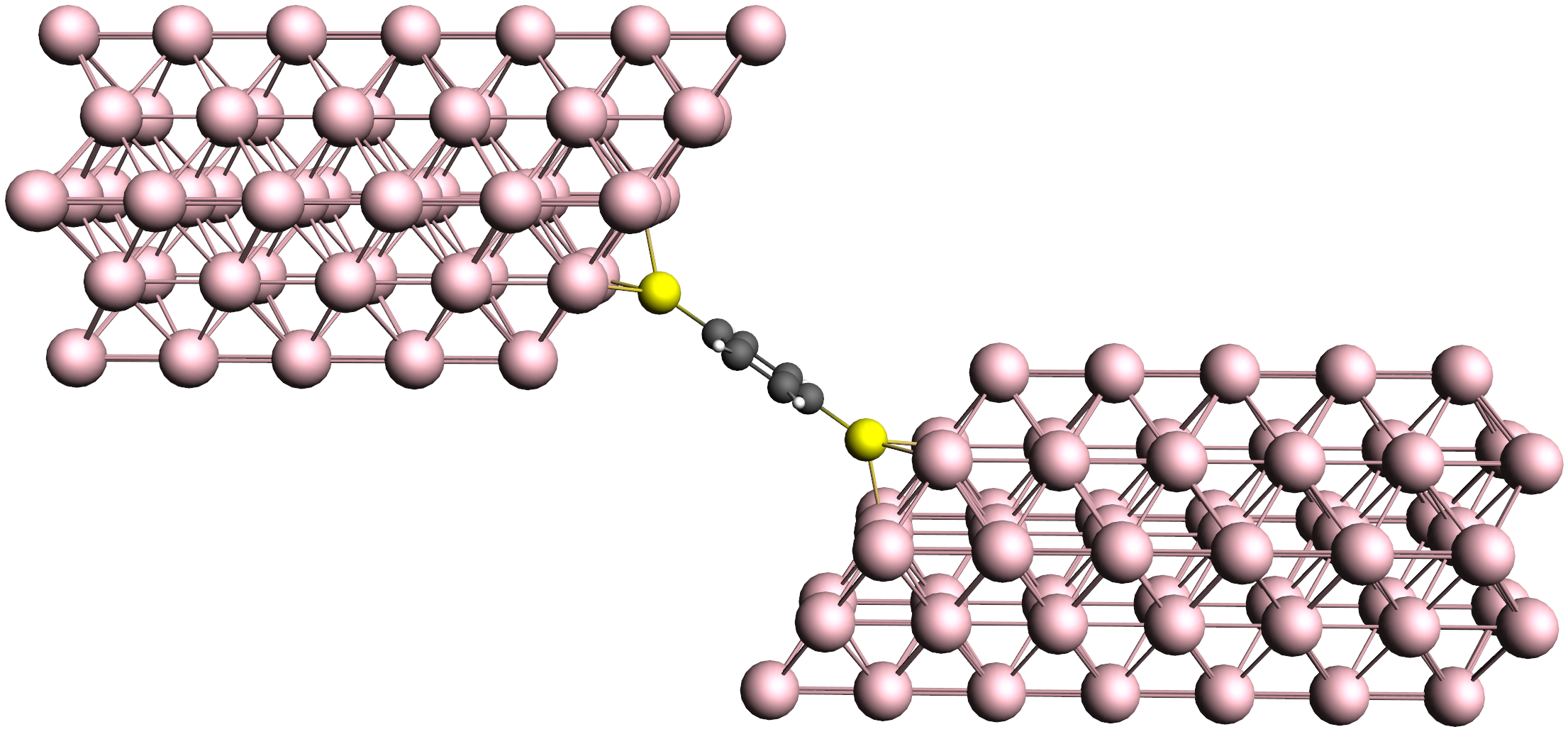}}\\
\subfloat[]{\includegraphics[width=.8\columnwidth]{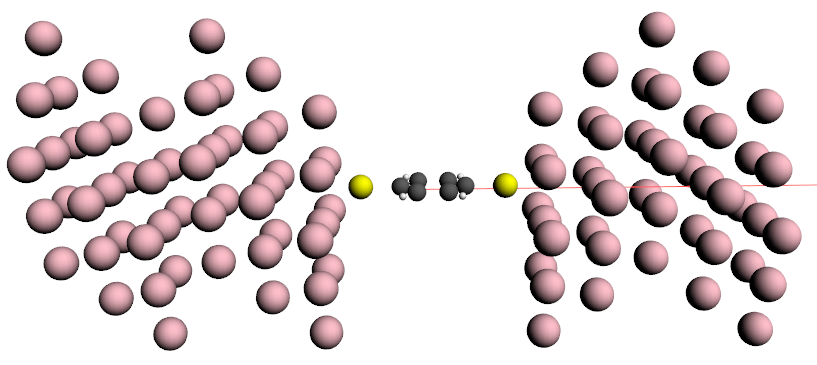}\label{fg:unslanted}}
\caption{Au contacts' geometry for {\bf a)} the $2\times 2$ surface case, {\bf b)} the $3\times 3$ surface case, both with hollow-site binding with 2.40 \AA\xspace Au-S distance. {\bf c)} An alternate $3\times 3$ surface used in calculations with bias, resp. with periodic boundary conditions, \emph{cf}. Fig.~\ref{fg:Au-BDT-Au_3x3_pbc_geom_II}. Calculations performed using the LDA functional and a SZ basis with 11 valence electrons.}\label{fg:Au.geoms}
\end{figure}

Before we proceed to transport calculations using ``bulky'' Au contacts, which will be used as a building block in all following sections, we first discuss their construction. We will emphasize, in particular, the difference between applying transverse \pbc and their omission. We begin the discussion with figures Fig.~\ref{fg:Au.Transmissions}-\ref{fg:Au.geoms}, where we illustrate 3 representative geometries one might use for the contacts, and the transmission through their bulk structure (\emph{without} molecules). These are all FCC stacked with a (111) face perpendicular to the axis, and we compare a transverse $2\times 2$--atom surface with the $3\times 3$--surface case, the latter both with and without {\pbc}. Calculations were performed with the LDA functional and a SZ basis with 11 valence electrons.

In appendix \ref{AlignmentTests} we show the convergence of the alignment shifts (Figs.~\ref{fg:Au.ShiftConvergence1}-~\ref{fg:Au.PDOS.Compares}), with the conclusion that in contrast to the case without \pbc, the shifts obtained in order to converge the extended-molecule structure with \pbc can be quite large.\footnote{We emphasize that this is an algorithmically-determined rather than an arbitrarily imposed shift, and that as we illustrate in Fig.~\ref{fg:Au.PDOS.Compares}, the PDOS indicate that the structure is correctly converged.}

It appears that for the case of \pbc, the bulk run in BAND is significantly offset in potential with respect to the alignment and transport runs. The strong difference with the alignment run argues for fixing the Fermi level via the correction $\dphi_0+\dphi_1$ (as opposed to simply neglecting a numerical error incorrectly assumed small). The continuing difference in the transport run further argues for a dynamic (runtime) correction $\dphi_0$, as implemented, rather than assuming a static correction $\dphi_0$ in transport.

In Fig.~\ref{fg:Au.Transmissions}, the typical transmission characteristics of these ``bulk'' junctions exhibit clear conductance plateaus, as we expect from what is essentially a ``bulky'' monatomic chain. We note that the number of channels found is similar near the Fermi level, while further away, in particular in the region from -8 to -2 eV, the structure \emph{without} \pbc is considerably noisier, and has fewer channels. 
This may be due in part to the electrons moving to the surface boundaries in the structure without \pbc, effectively reducing the number of transport channels. 

Having examined the alignment, convergence and bulk structure of the extended-molecules in different cases, we now proceed to single-molecule calculations. We will mainly use $3\times 3$ surfaces of Au, both with and without \pbc.

\subsection{Benzenedithiol Junctions}

\begin{figure}
\subfloat[]{\includegraphics[width=.8\columnwidth]{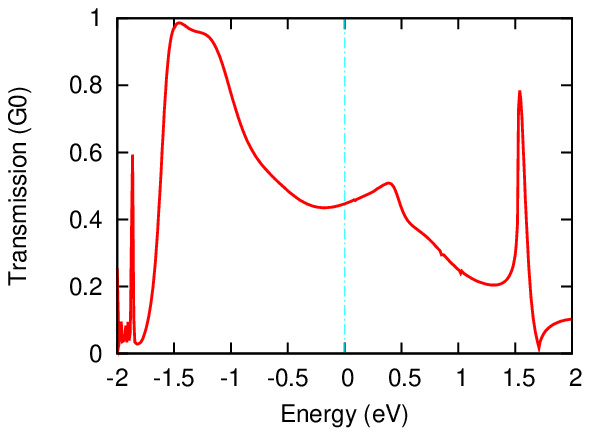}\label{fg:2x2}}\\
\subfloat[]{\includegraphics[width=.8\columnwidth]{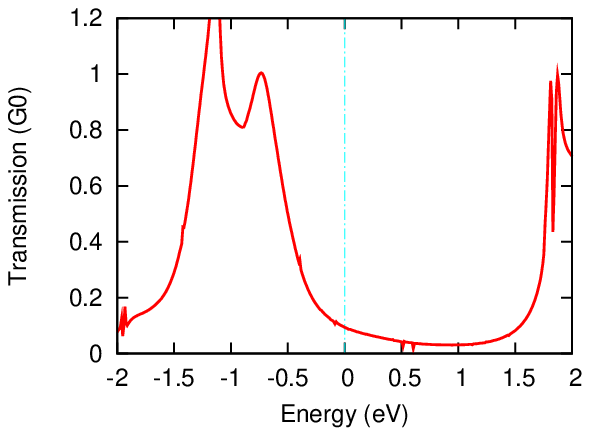}\label{fg:3x3}}\\

\caption{Transmission through the Au--BDT--Au junction, using {\bf a)} $2\times 2$ and {\bf b)} $3\times 3$ 
Au contacts without periodic boundary conditions. We see that the transport gap opens, becoming recognizably linked to the smooth structure visible below in Fig.~\ref{fg:Au-BDT-Au_3x3_pbc_geom_II} for the case of \pbc. The main peaks map between the two sets of calculations, and we illustrate the orbitals that play the dominant role for the $3\times 3$ case in Fig.~\ref{fg:AuBDT.orbitals} below.}\label{fg:AuBDT.Transmission1}
\end{figure}

We now consider the well-studied ``test case'' of a gold-benzenedithiol-gold (Au--BDT--Au) junction.\cite{Xue2003a,Xue2003b,Stokbro2003,Stokbro2003a,Nara2004,Pontes2006} The contacts are designed in the same way as in the previous section, with Au (111) faces consisting of $2\times 2$ and $3\times 3$ atoms,
shown in Fig.~\ref{fg:Au.geoms}. Except where stated otherwise, we perform transport calculations in our code and gas-phase calculations in ADF using the LDA functional with a SZ basis on the Au contacts and a TZP basis on the molecule. In Fig.~\ref{fg:AuBDT.Transmission1} we present the results of our calculations without \pbc. 

We first note that 
on increasing the size of the surface perpendicular to transport, there is a
relatively quick convergence to a recognizable result with a broad HOMO-like peak below $\epsilon_f$, followed by a 2--3 eV  low-conductance gap separating it from a LUMO peak beyond the gap, around 2 eV. This confirms that the size of contacts does matter to the calculation, though the major features already become established for modest contact sizes.

\begin{figure}
\subfloat[LUMO+1, acts as LUMO]{\includegraphics[width=.5\columnwidth]{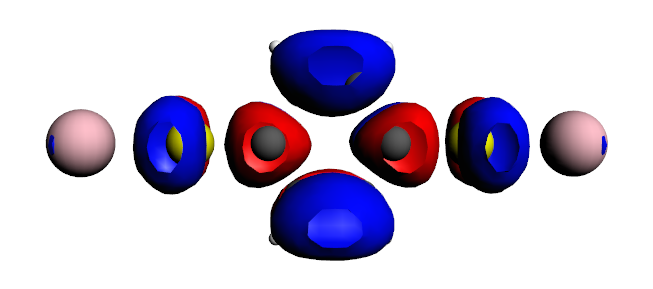}}
\subfloat[LUMO, does not couple]{\includegraphics[width=.5\columnwidth]{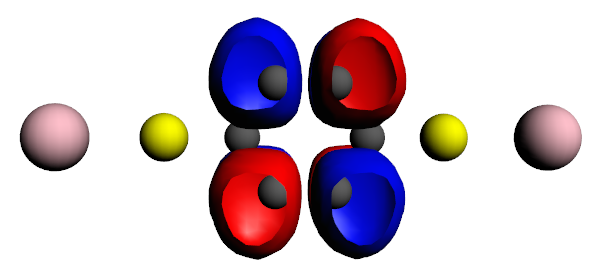}}\\
\subfloat[HOMO]{\includegraphics[width=.5\columnwidth]{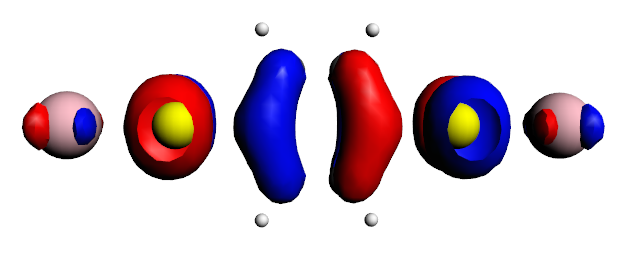}}
\subfloat[H$_\text{A}$: apparent HOMO state on fragment]{\includegraphics[width=.5\columnwidth]{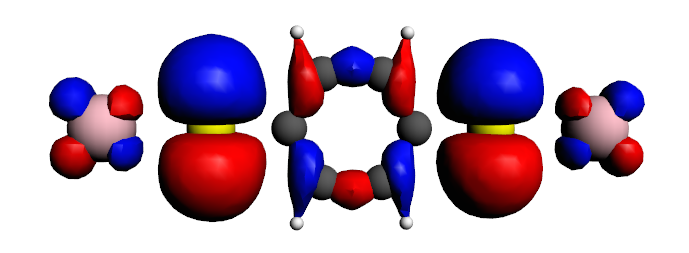}\label{fg:appH}}\\
\subfloat[H$_\text{B}$: apparent HOMO-1 state on fragment]{\includegraphics[width=.5\columnwidth]{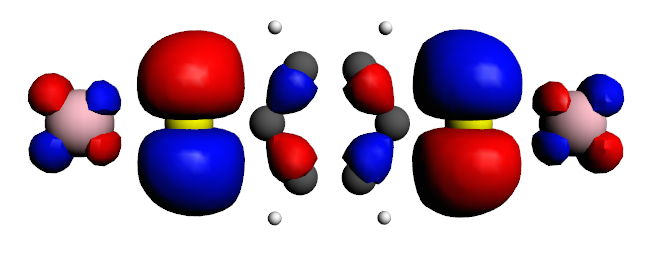}\label{fg:appH-1}}
\subfloat[HOMO-1]{\includegraphics[width=.5\columnwidth]{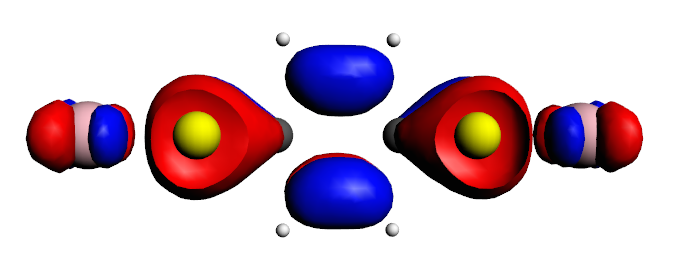}}\\
\subfloat[HOMO-2, does not couple]{\includegraphics[width=.5\columnwidth]{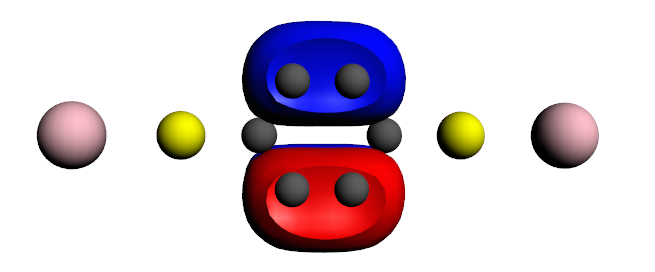}}
\subfloat[HOMO-3, couples below -2eV]{\includegraphics[width=.5\columnwidth]{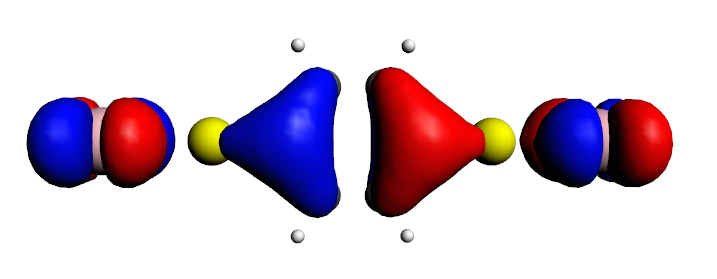}}
\caption{Transport-coupled orbitals of the BDT junction,
ordered by decreasing chemical potential. {\bf a)--c)} and {\bf f)--h)} are the fragment's orbitals nearest the Fermi level, which are labeled by correspondence to their gas-phase counterparts. {\bf d)--e)} are two examples of intermediate states that are the present in the fragment but not the neutral BDT molecule, and survive the thiolate coupling to Au adatoms instead of the terminal -SH bond.}\label{fg:AuBDT.orbitals}
\end{figure}

\begin{figure}
\subfloat[]{\includegraphics[width=\columnwidth]{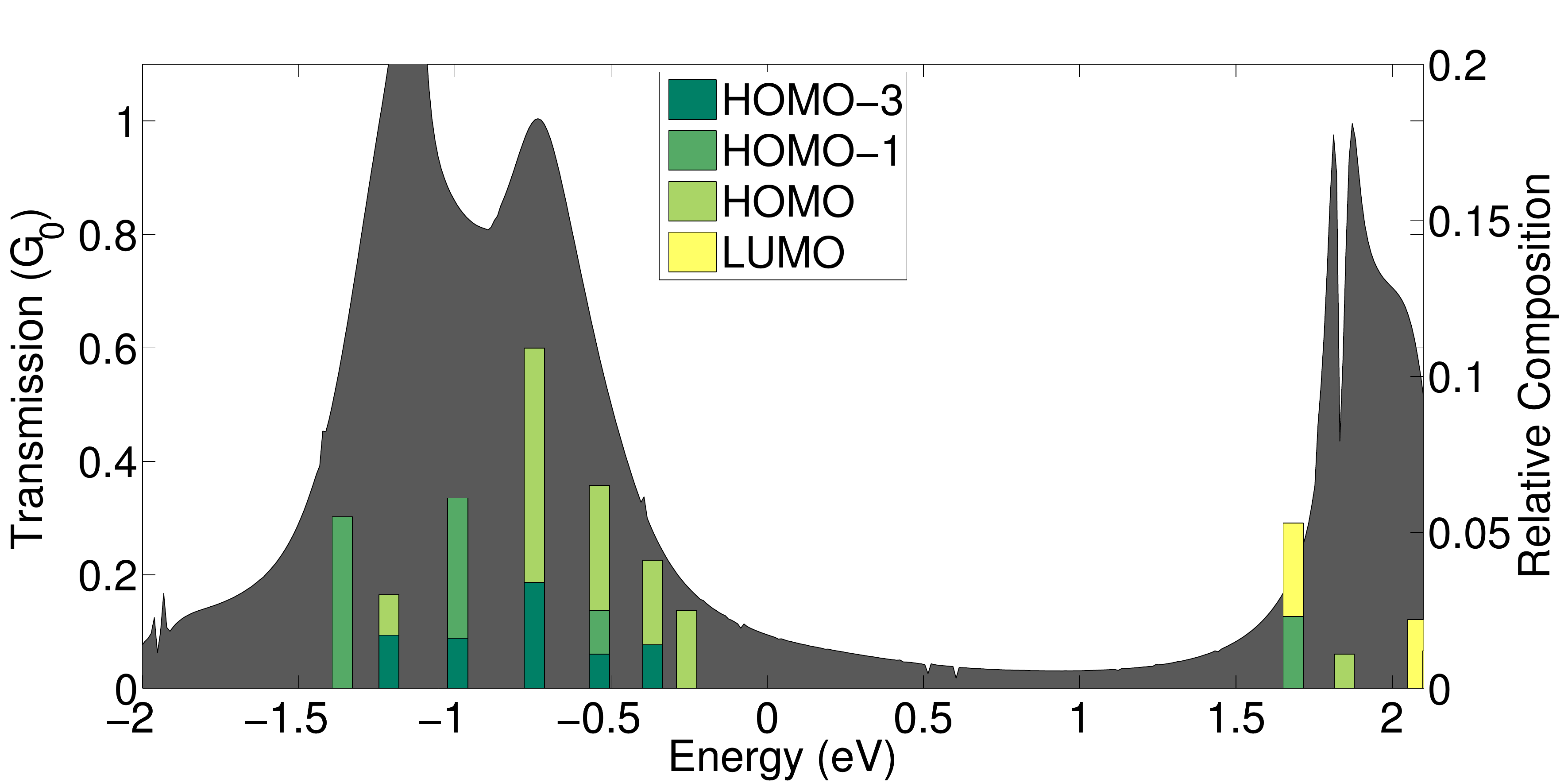}}\\
\subfloat[]{\includegraphics[width=\columnwidth]{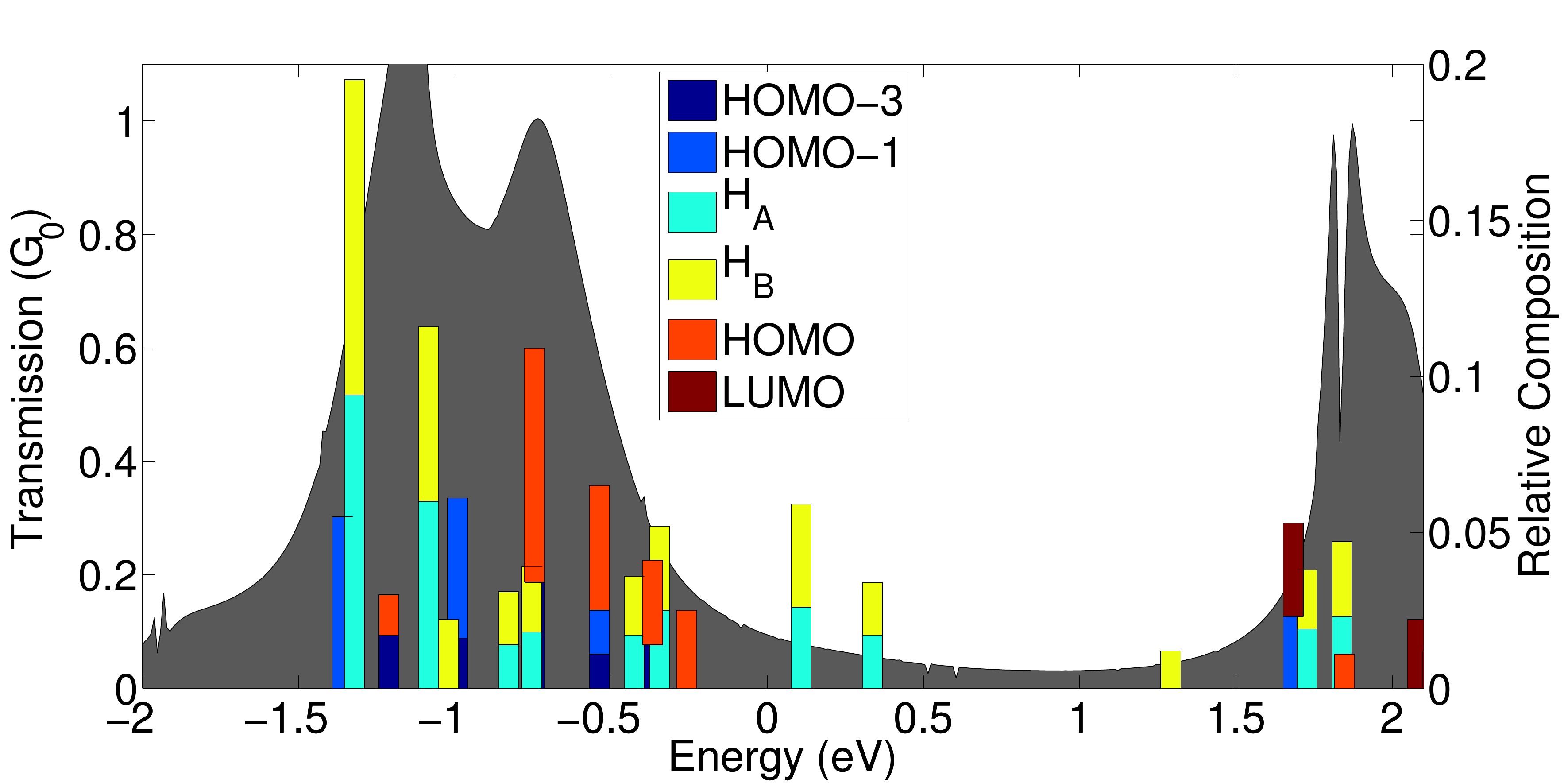}}
\caption{Transmission and peak decompositions for the $3\times 3$ Au--BDT--Au junction. Distance between clearest frontier peaks at -0.26 eV and 1.68 eV suggests an effective gap of roughly 1.93 eV. (a) Illustrates the decomposition onto the main gas-phase orbitals, while (b) also includes the H$_\text{A}$ and H$_\text{B}$ orbitals of Fig.~\ref{fg:appH} and Fig.~\ref{fg:appH-1}.\label{fg:bdt.peaks}}
\end{figure}

Fig.~\ref{fg:AuBDT.orbitals} shows the main orbitals derived from the BDT molecule, labeled by their correspondence to the gas-phase orbitals. We use these in order to construct Fig.~\ref{fg:bdt.peaks}, showing the compositions of the peaks in the transmission through the Au--BDT--Au junction near $\epsilon_f$. 

We have determined these by using a fragment-decomposition technique, outlined in \ref{decompositions}, in which we project the eigenstates of the transport calculation onto the orbitals of a molecular fragment. As a fragment we use the gas-phase BDT molecule geometry with thiolate bonds to a single Au atom on each side (outermost H's of the gas-phase BDT are removed, leaving an Au--S bond), as expected for the preferred bonding of a molecule ending on a thiol group, \emph{e.g.} BDT, to an Au surface.\cite{Love2005,Hoft2006,Romaner2006}

We can identify the orbitals of the fragment with those of either the BDT molecule or radical, and this flexibility also provides extra information on the complicated orbital compositions which we find in the junction. We find that adding the Au adatoms induces the formation of hybrid Au--BDT--Au states which may couple well in transport, labeled H$_\text{A}$ and H$_\text{B}$ in Fig.~\ref{fg:AuBDT.orbitals}, to reflect their energy ordering as the ``apparent'' HOMO and HOMO-1 states on the fragment. We discuss their role in more detail later. The rest of the states in Fig.~\ref{fg:AuBDT.orbitals} are labeled by their correspondence to the orbitals of gas-phase BDT. We now focus on the geometry of Fig.~\ref{fg:3x3} in particular, as we later use it to model OPE-2 and OPE-3 as well.

We find that the broad HOMO-resonance appears to be mainly composed of 2 separate peaks, which we identify by our decomposition analysis as the HOMO and HOMO-1 peaks of the gas-phase BDT molecule, with a bit of HOMO-3 playing a role as well. H$_\text{A}$ and H$_\text{B}$ also appear here. In the fragment they are split by about 120 meV, and appear as a result of hybridization with Au, or more generally, for collinear termination on the thiol (\emph{i.e.} they also occur when the -SH bond in gas-phase BDT is collinear). This suggests that they may represent a bonding/anti-bonding pair which interferes nearly perfectly destructively, and as a result does not contribute much to transport.\cite{Hoffmann1971,Solomon2010} In this context we further note that despite the $\sim 120$ meV splitting, they indeed consistently appear in roughly equal measure in each contribution in Fig.~\ref{fg:bdt.peaks}b, once coupled to Au.
The LUMO and HOMO-2 states, by contrast, are sufficiently localized to the center of the molecule that do not couple in transport; rather, for the unoccupied states it is the LUMO+1 which appears as the lowest unoccupied transport peak above the transport gap. We return to this point in discussing the OPE-series in the next section. 
We do find the HOMO-2 peak present in our decomposition around -2.45 eV, and as a considerably more pure state, consistent with the lack of coupling to the Au contacts. By contrast, the HOMO-3 has an orbital structure which suggests coupling in transport, and it is present, mixed with the HOMO and HOMO-1 states, in Fig.~\ref{fg:bdt.peaks}. This may suggest an analogy to the HOMO-2 states of the OPE-series.\\

Finally, we also remark on the presence of some small discontinuities in the transmissions in Fig.~\ref{fg:AuBDT.Transmission1}, in the $3\times 3$ case near \emph{e.g.} -0.4 eV, -0.1 eV and 0.5 eV. We have investigated these using the projected DOS on the molecule proper, extended molecule and deep contacts. We find that these are not numerical artifacts, but rather are related to effects in the potential in the contacts. While these may be relevant for very sharp, needle-like contacts, they would probably not play a role for relatively large, bulk-like electrodes. \\

While we remarked that the results of Fig.~\ref{fg:AuBDT.Transmission1} indicate a gradual convergence towards a ``bulk'' face result (the single benzenedithiol atom coupled to an infinite plane of Au on each side), they should be distinguished from the ``classic'' Au--BDT--Au junction results calculated using DFT+NEGF in the literature.\cite{Stokbro2003,Xue2003a,Xue2003b} 
The difference is the absence of periodic boundary conditions here, a relatively important modeling decision not typically discussed in the early literature, in large part because few codes allow for explicitly breaking transverse periodicity. We illustrate this difference explicitly in Fig.~\ref{fg:AuBDT.Transmission2} below, which should be compared with Fig.~\ref{fg:3x3}.
The implications of the use of periodic boundary conditions for a junction with such a small face evident in Fig.~\ref{fg:Au-BDT-Au_3x3_pbc_geom_II}, where we see that the model in this case is qualitatively more similar to a self-assembled monolayer than to a true single-molecule configuration. Conversely, the geometry without \pbc are particularly useful for modeling the small, needle-shaped contacts used in break-junction experiments. This difference has consequences for the conductance of the system,\cite{Liu2005} but for larger inter-molecular separation this need not be an issue per se, as long as the system being modeled is not actually needle-like in geometry.\\

\begin{figure}
\subfloat[]{\label{fg:AuBDT.Transmission2}\includegraphics[width=.85\columnwidth]{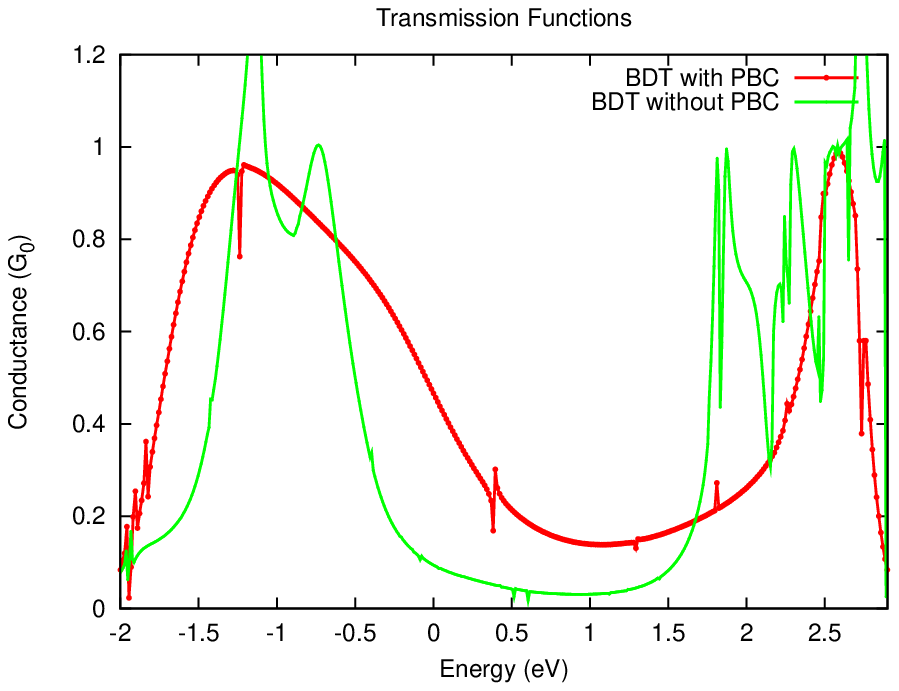} }\\
\subfloat[]{\label{fg:Au-BDT-Au_3x3_pbc_geom_II}\includegraphics[width=.8\columnwidth]{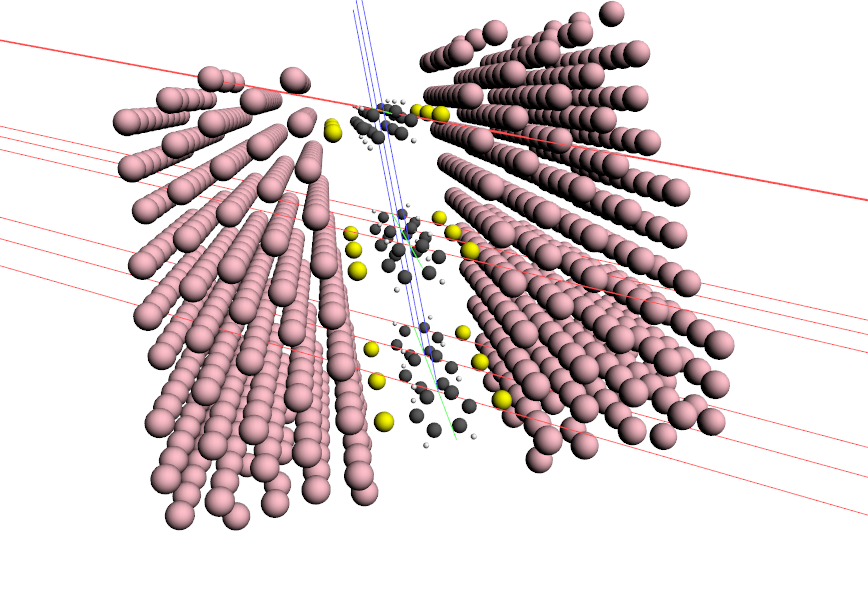}}\caption{{\bf a)} Transmission through the Au--BDT--Au junction with \pbc, compared with the previous result without \pbc, 
Fig.~\ref{fg:AuBDT.Transmission1}. {\bf b)} Implied geometry of the model $3\times 3$ Au--BDT--Au junction, using periodic boundary conditions. Overall, we observe a further opening of the gap and a broadening of the HOMO-like peaks, which themselves are reduced in magnitude back to $G_0$ transmission.}
\end{figure}

Finally, for Fig.~\ref{fg:AuBDT.Transmission2} we find the same general picture in terms of peaks (compositions not illustrated). 
 The predominance of the HOMO, HOMO-1 and their ``apparent'' counterpart states in transport is, moreover, in excellent agreement with the findings of Stokbro and others.\cite{Stokbro2003,Kondo2006}

Our work indicates that the convergence towards this bulk result is primarily dependent on having a sufficient number of transport channels available in the contacts to couple to, as in Fig.~\ref{fg:Au.Transmissions}. This has implications for contacts which do not couple to as many channels, where a broad peak from the strongly-coupled regime may break apart into a number of narrower (less strongly-coupled) resonances, though still enabling transport. A noteworthy feature is that the transmission \emph{at} the Fermi level\ $T(\epsilon_f)$ may be significantly reduced, bringing the value closer to the experimental one,\cite{Ulrich2006,Hybertsen2008} where one typically would expect to be in the weakly-- rather than strongly-coupled regime.

\subsection{OPE--series Junctions}

\begin{figure}
\subfloat[Au--OPE-2--Au Junction Geometry]{ \includegraphics[width=\columnwidth]{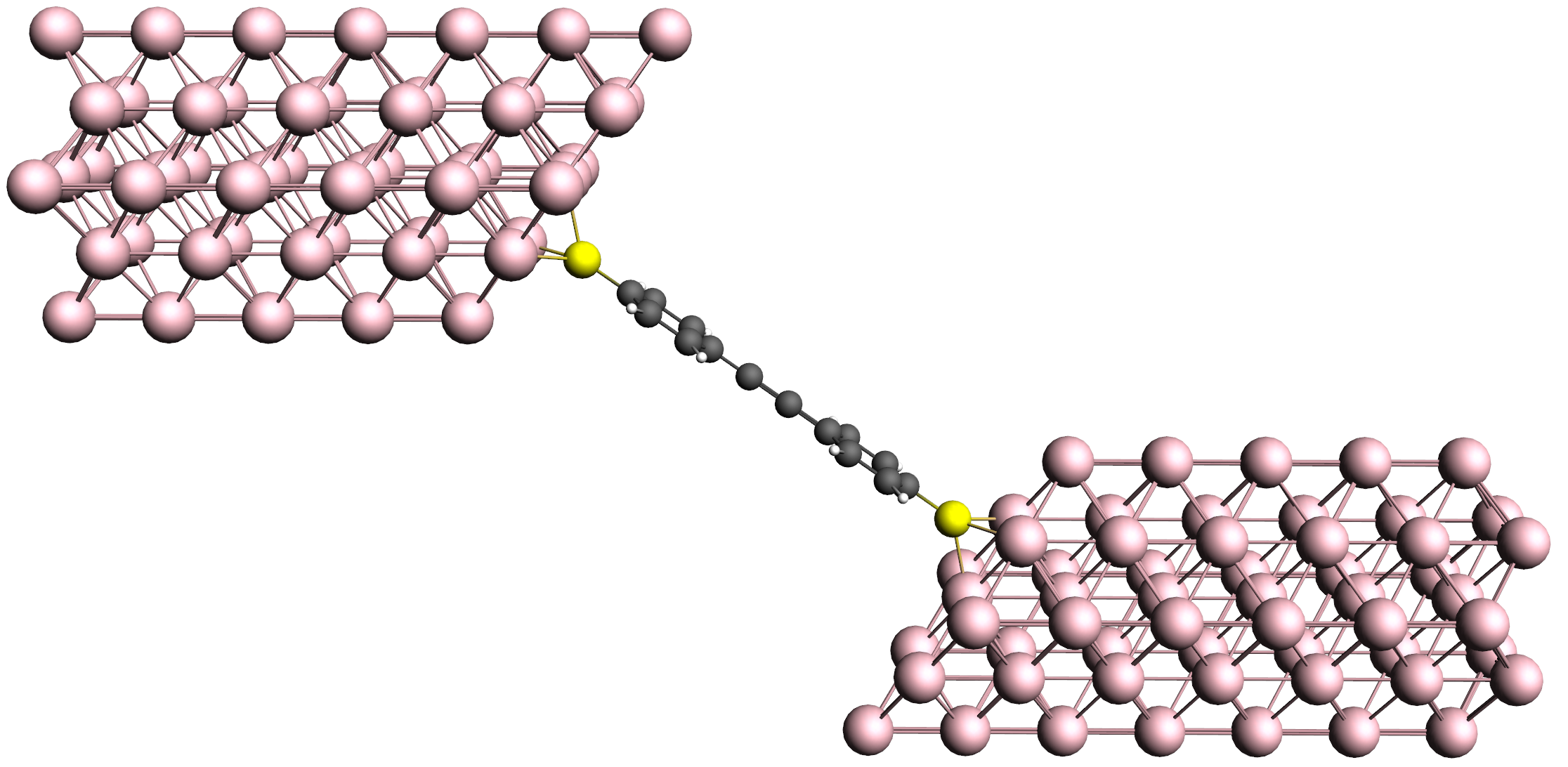}\label{fg:OPE2.geom} }\\
\subfloat[Au--OPE-3--Au Junction Geometry]{ \includegraphics[width=\columnwidth]{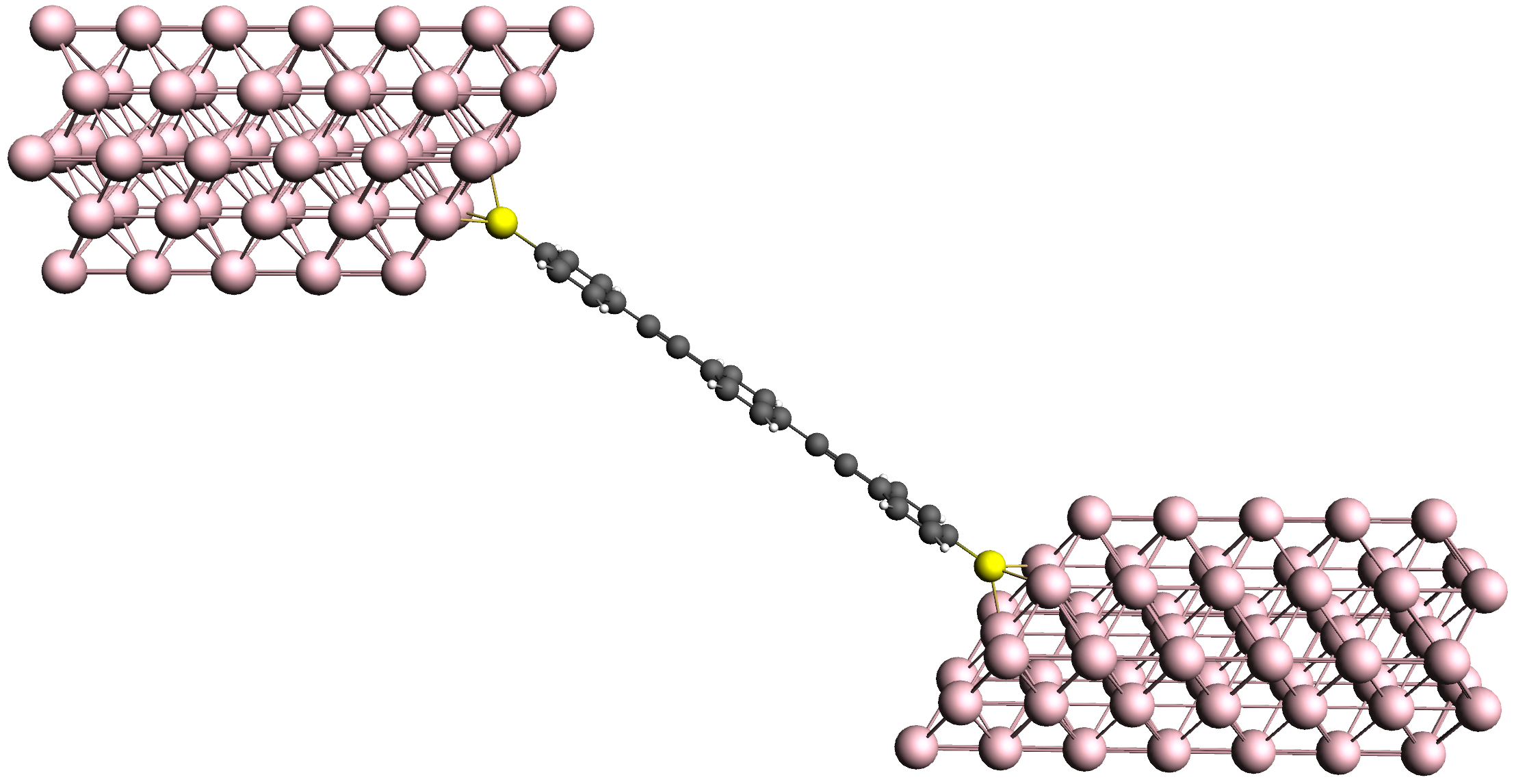}\label{fg:OPE3.geom} }
\caption{Geometry of $3\times 3$ atom (111) surface Au--OPE-2--Au and Au-OPE-3--Au junctions, both without periodic boundary conditions. Hollow-site binding with 2.40 \AA\xspace Au-S distance, compare Fig.~\ref{fg:Au.geoms}.}\label{fg:OPE.geom}
\end{figure}

We now proceed by considering the first two of the thiol-anchored oligophenylene-ethynylene family of molecules. Except where stated otherwise, we again perform transport calculations in our code and gas-phase calculations in ADF using the LDA functional with a SZ basis on the Au contacts and a TZP basis on the molecule. We show the results of modeling OPE-2 and OPE-3 single-molecule junctions, with 2 and 3 phenyl rings respectively; the junction geometries are illustrated in Fig.~\ref{fg:OPE.geom}a--b respectively. These calculations use the same contacts as with BDT, and therefore have a common and well-determined Fermi level $\epsilon_f$, as discussed previously. These molecules have also been studied experimentally as promising benchmark systems in molecular electronics.\cite{Reichert2002,Mayor2003,Xiao2005,Huber2008}

\begin{figure}
\includegraphics[width=\columnwidth]{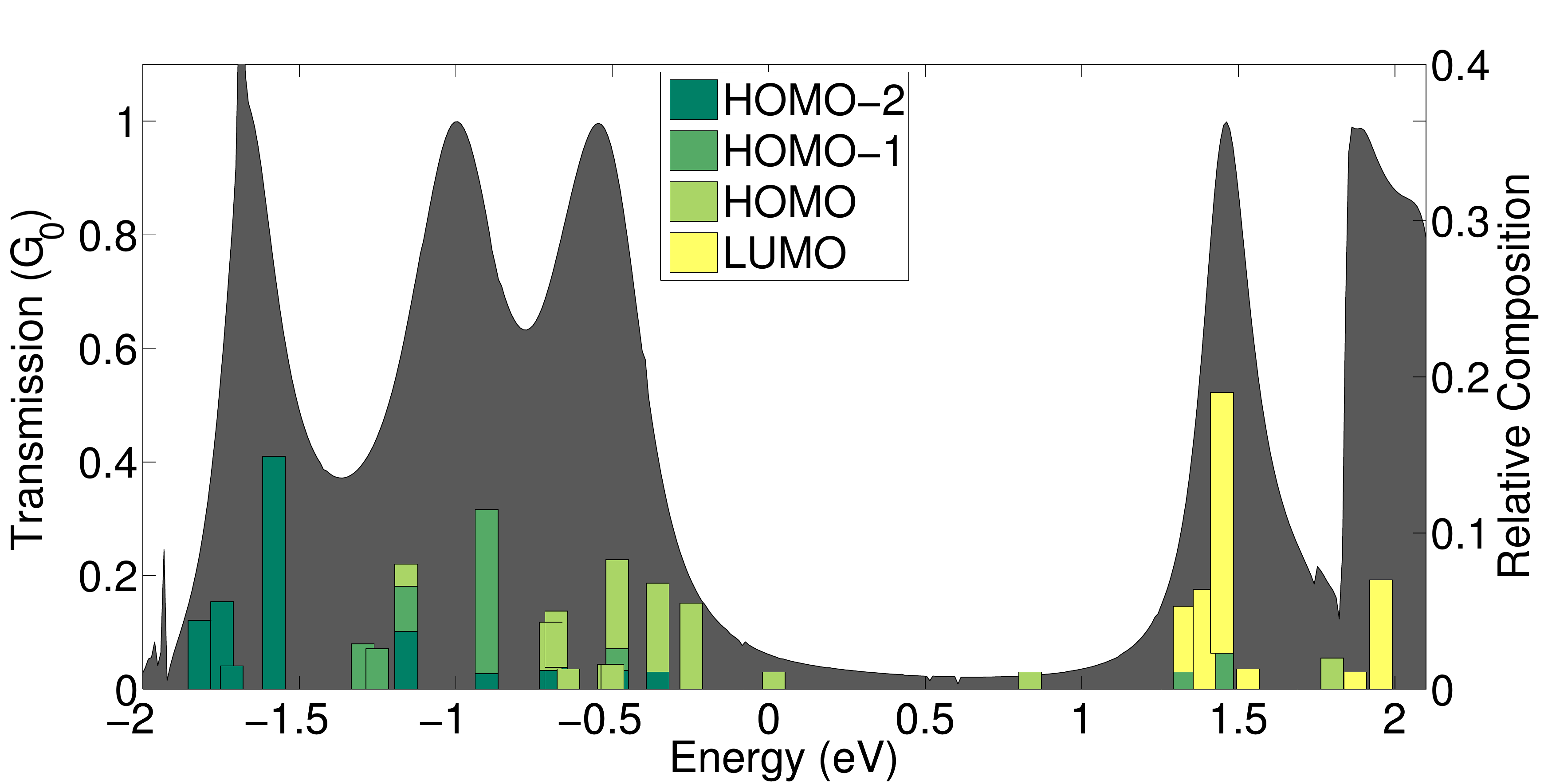}
\caption{Peak compositions near $\epsilon_f$ for the Au--OPE-2--Au junction. Distance between clearest frontier peaks at -0.25 eV and 1.33 eV suggests an effective gap of roughly 1.6 eV.}\label{fg:OPE2.transmission}
\end{figure}

\begin{figure}
\subfloat[LUMO, see Fig.~\ref{fg:AuBDT.orbitals}]{\includegraphics[width=.5\columnwidth]{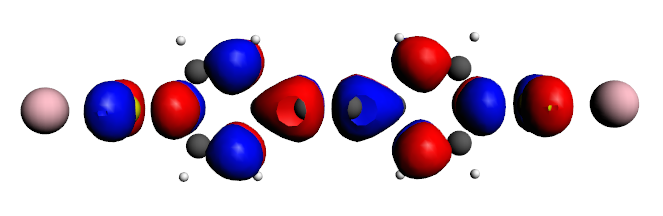}}
\subfloat[HOMO]{\includegraphics[width=.5\columnwidth]{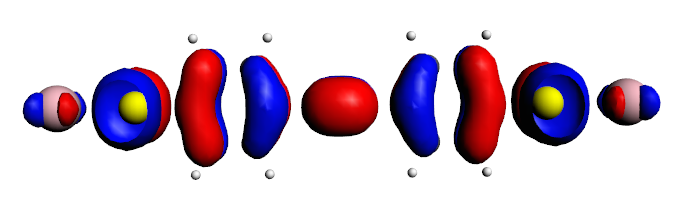}}\\
\subfloat[H$_\text{A}$: apparent HOMO state on fragment]{\includegraphics[width=.5\columnwidth]{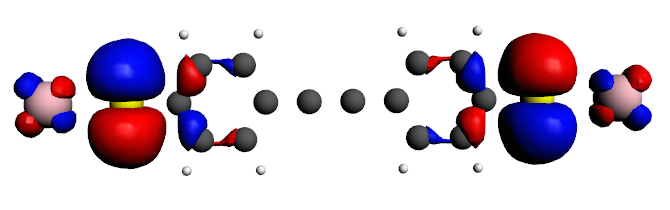}}
\subfloat[H$_\text{B}$: apparent HOMO-1 state on fragment]{\includegraphics[width=.5\columnwidth]{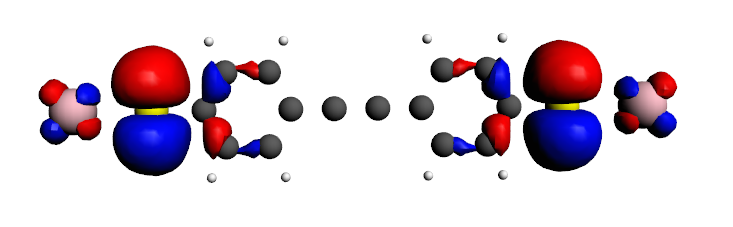}}\\
\subfloat[HOMO-1]{\includegraphics[width=.5\columnwidth]{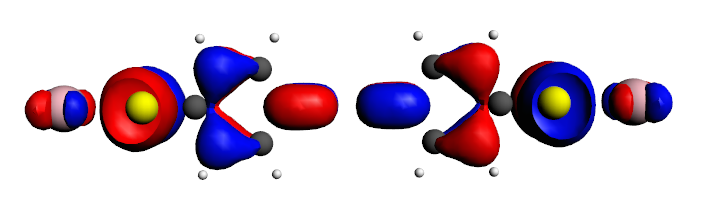}}
\subfloat[HOMO-2]{\includegraphics[width=.5\columnwidth]{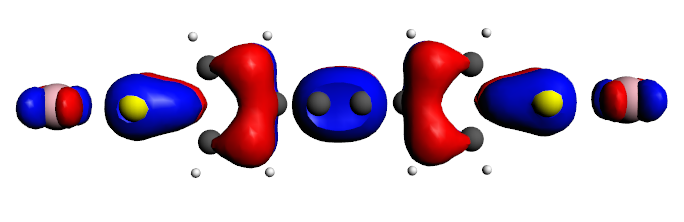}}\\
\caption{{\bf a)--f)} Fragment transport-coupled orbitals nearest the Fermi level (compare Fig.~\ref{fg:AuBDT.orbitals}). We note again H$_\text{A}$ and H$_\text{B}$, the ``apparent'' HOMO and HOMO-1 states, which are now more localized at the interface and whose wavefunctions do not extend all the way across the molecule.}\label{fg:OPE2.Combi}
\end{figure}

The transmission through a junction composed of OPE-2 coupled to $3\times 3$ atom Au (111) face contacts is illustrated in Fig.~\ref{fg:OPE2.transmission}, where the peak decompositions are constructed as outlined above for BDT. The fragment states to which the figure refers are illustrated in Fig.~\ref{fg:OPE2.Combi}, again labeled according to the gas-phase OPE-2 molecule's orbitals. In combination with these orbitals, we analyze the nature of transport, and the relation to the single-phenyl BDT system discussed previously.\\

The electronic structure near $\epsilon_f$ immediately recalls the results for the BDT junction. We again find a broad resonance, now split over three peaks below the Fermi level which are identified with the HOMO-2, --1 and HOMO states. The broad peaks are further composed of mixtures with the H$_\text{A}$ and H$_\text{B}$ states of the Au-OPE fragments and Au-derived states that are mainly localized at the Au-S bond. Beyond the transport gap, at 1.5 eV, we again find a peak which is identified predominantly with the gas-phase LUMO. This is a state which on the OPE-2 (and OPE-3) junction has an orbital symmetry that immediately recalls the LUMO+1 of BDT. The LUMO+1 states on these two molecules, conversely, recall the LUMO of BDT, and do not play a strong role in transport due to the localization of electrons away from the contacts.\\

\begin{figure}
\includegraphics[width=\columnwidth]{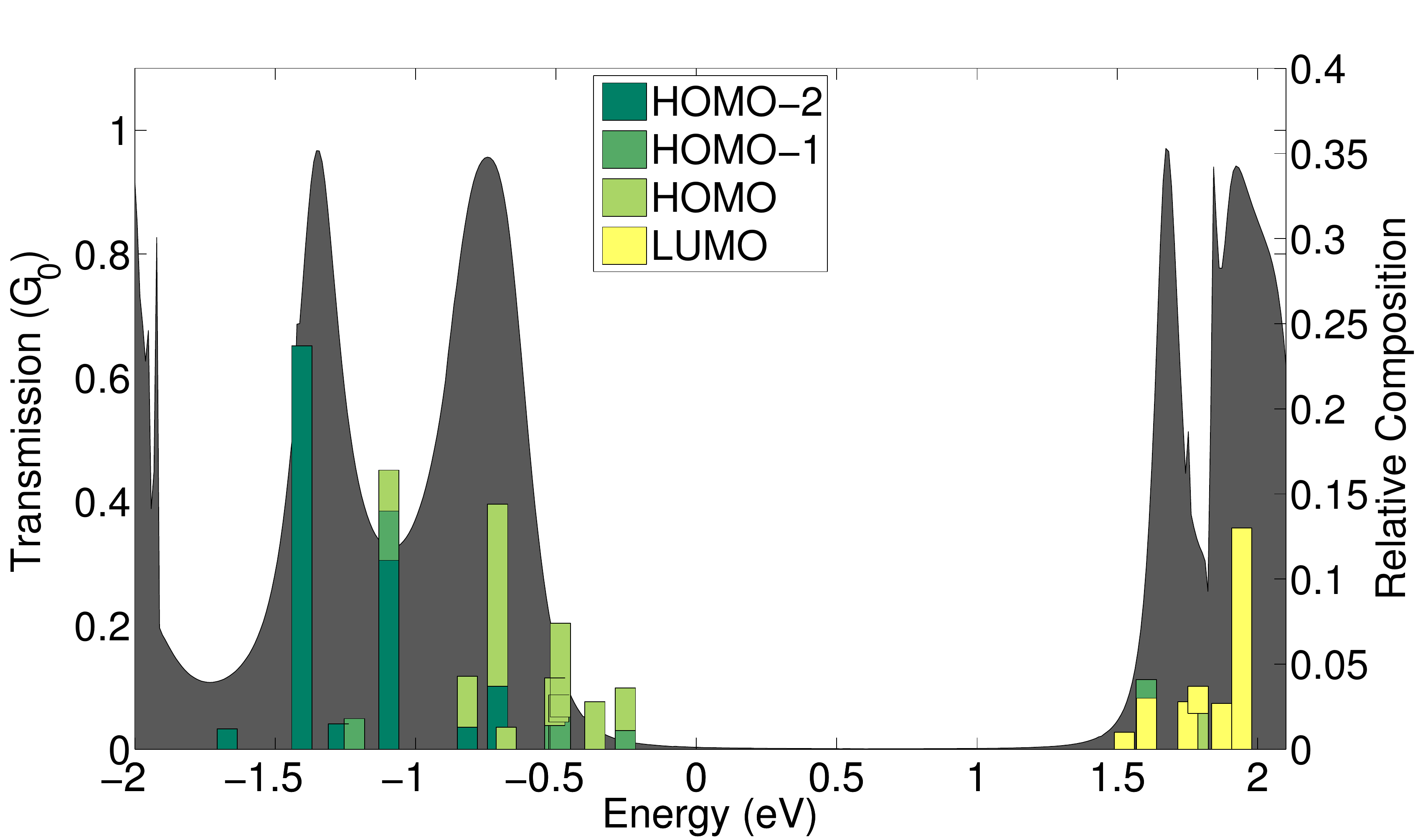}
\caption{Peak compositions near $\epsilon_f$  for the Au--OPE-3--Au junction. Distance between clearest frontier peaks at -0.25 eV and 1.52 eV suggests an effective gap of roughly 1.8 eV.}\label{fg:OPE3.transmission}
\end{figure}

\begin{figure}
\subfloat[LUMO]{\includegraphics[width=.5\columnwidth]{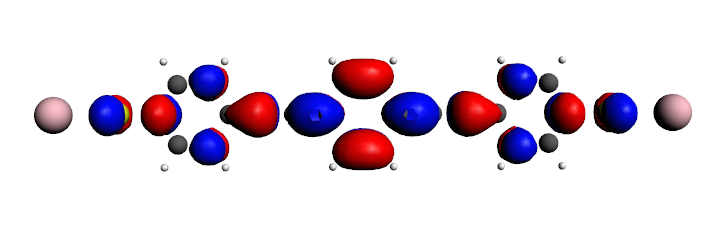}}
\subfloat[HOMO]{\includegraphics[width=.5\columnwidth]{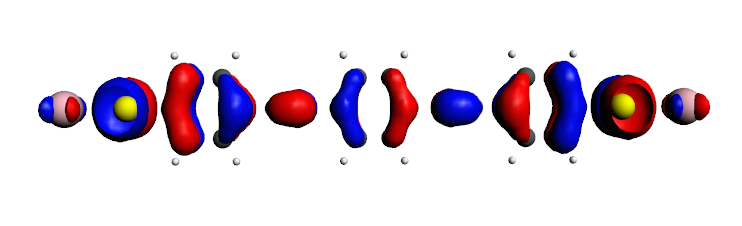}}\\
\subfloat[H$_\text{A}$: apparent HOMO state on fragment]{\includegraphics[width=.5\columnwidth]{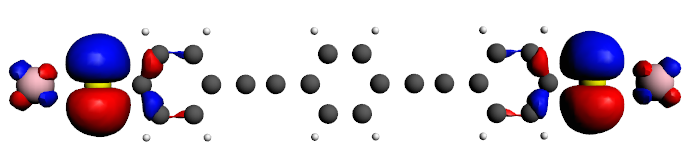}}
\subfloat[H$_\text{B}$: apparent HOMO-1 state on fragment]{\includegraphics[width=.5\columnwidth]{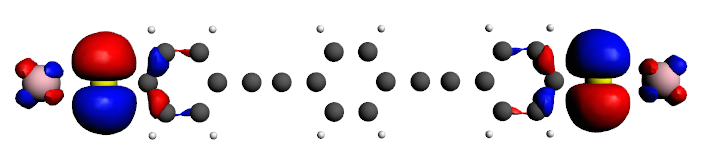}}\\
\subfloat[HOMO-1]{\includegraphics[width=.5\columnwidth]{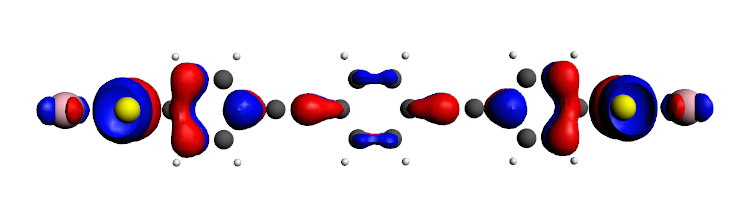}}
\subfloat[HOMO-2]{\includegraphics[width=.5\columnwidth]{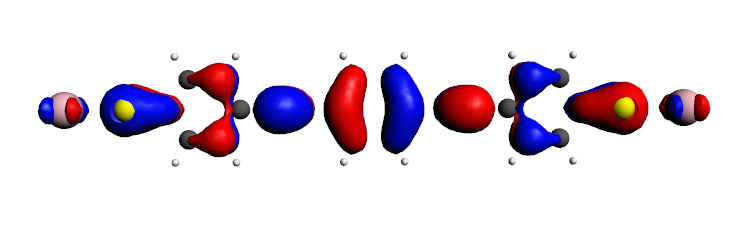}}\\
\caption{{\bf a)--f)} Fragment transport-coupled orbitals nearest the Fermi level (compare figures Fig.~\ref{fg:AuBDT.orbitals} and Fig.~\ref{fg:OPE2.Combi}). We again note H$_\text{A}$ and H$_\text{B}$, again the ``apparent'' HOMO and HOMO-1 states, and analogues of Fig.~\ref{fg:OPE2.Combi} {\bf c)--d)} which are again localized mostly near the S atom, and as with OPE-2 have wavefunctions which do not extend all the way across the molecule.}\label{fg:OPE3.Combi}
\end{figure}

\begin{figure}
 \includegraphics[width=\columnwidth]{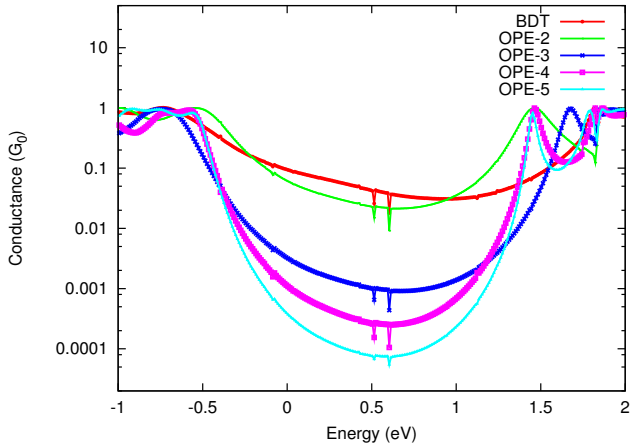}
\caption{Comparison of transmissions on log-scale showing evolution of main features near the Fermi level as molecule in junction is varied from 1 to 3 phenyl rings. Opening of the transport gap is visible, as well as narrowing of the peaks near $\epsilon_f$ and a general spectral shift backwards of the HOMO-like peaks.}\label{fg:MultiRingComparison}
\end{figure}

For OPE-3, the transmission is illustrated in Fig.~\ref{fg:OPE3.transmission} together with the peak decompositions. The orbitals referred to are illustrated in Fig.~\ref{fg:OPE3.Combi}, which further confirms the picture nature of transport in this family.  When comparing with the Au--BDT--Au and Au--OPE-2--Au junctions' results, it appears that the gap slightly reopens. This, however, still appears to be part of a progression towards a smaller transport gap for the longer molecules, following the trend towards smaller HOMO-LUMO gap. This is illustrated in the log-scale plot of Fig.~\ref{fg:MultiRingComparison}, where we show the transmission for OPE-4 and OPE-5 junctions as well.

Again, it is the gas-phase HOMO which dominates the conductance near the Fermi level, with the HOMO-1 and HOMO-2 below it composing the lower-lying peaks. The LUMO dominates beyond the transport gap around 1.5--2 eV. The LUMO of OPE-3 again recalls the LUMO+1 orbital of BDT found to compose the ``effective LUMO'' peak there, and so we confirm the role of the orbital symmetries between the respective ``LUMO,'' HOMO, HOMO-1 and HOMO-2 states for each of the three molecules.

The H$_\text{A}$ and H$_\text{B}$ states, see Fig.~\ref{fg:OPE3.Combi}, are likewise present in all 3 junctions, but as the molecule becomes larger, they become more localized in nature than conjugated, and so should play progressively less of a role in transport through the longer molecules of the family. However, in practice, we find them present in roughly similar proportion to the HOMO and HOMO-1 states in the broad peaks right below $\epsilon_f$ in our decompositions for all 3 systems, as well as in the low-conductance transport gap, which suggests that, as argued above, these two states may not really be contributing to transport.

The big picture then, appears to be a slightly erratic change of the gap, which seems to converge from the third or fourth member of the OPE$-n$ family on, narrowing the gap as the number of phenyl-rings in the junction molecule increases. As pointed out by Ke \emph{et al.},\cite{Ke2007} this should converge to a gap similar to the infinite-OPE-chain band gap of $\sim 1.7$ eV,\footnote{The value of $\Delta\sim 1.72$ eV calculated by us using the BAND code, with 1D periodicity using LDA with a TZP basis set to match the transport calculations; compare $\sim 1.5$ eV as calculated using GGA with unspecified basis set, by Ke \emph{et al.}} dominated by the molecular orbitals as the influence of the contacts begins to decrease with wire length. We remark that using carbon nanotube contacts, they observe similarly erratic behavior where the gap-size is concerned. Finally, we remark on the clear trend towards steadily lower values of $T(\epsilon_f)$ as the molecular wire length increases, also visible in Fig.~\ref{fg:MultiRingComparison}.

\subsection{Finite Bias Calculations}\label{bias}

\begin{figure}
\subfloat[]{\includegraphics[width=.95\columnwidth]{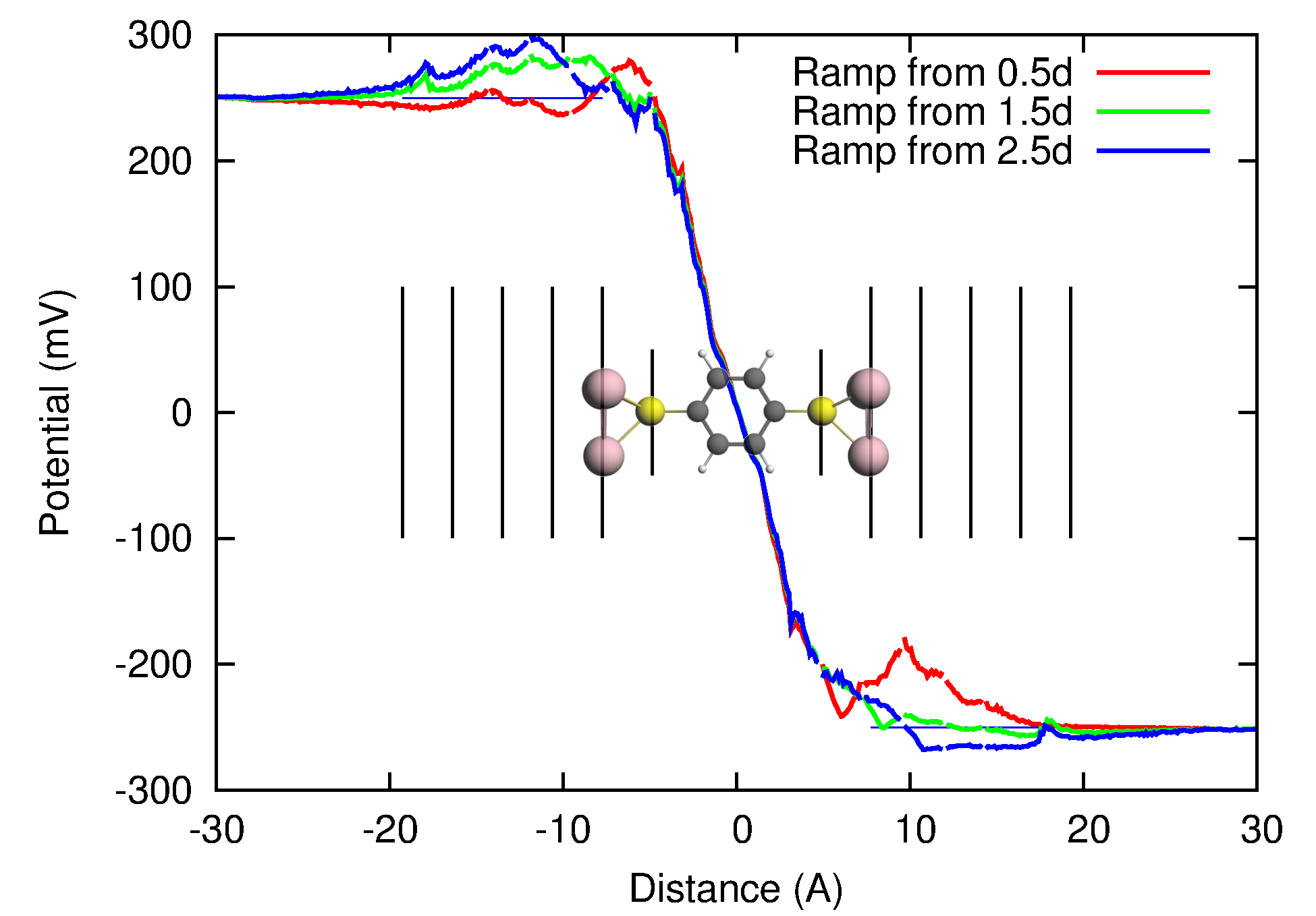} \label{fg:BDT_bias_profile}}\\
\subfloat[]{\includegraphics[width=.8\columnwidth]{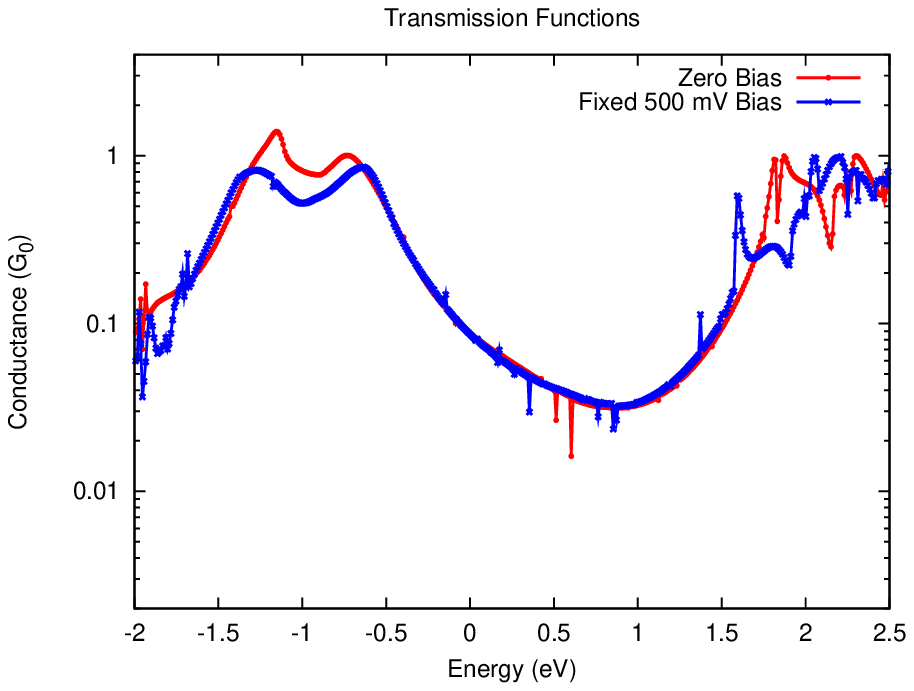} \label{fg:AuBDT.bias}}
\caption{{\bf a)} Potential drop with fixed ramps starting at $0.5d, 1.5d$, and $2.5d$ from the innermost Au layer of the device region, where $d$=2.88\AA\xspace the interlayer spacing. $500$mV bias is applied, results are relative to a zero-bias self-consistent calculation. Molecule with nearest Au neighbors indicated, together with vertical black lines showing position of subsequent layers of Au. Note the decreased slope of the potential drop over the benzene ring, as compared with the Au-S bond region. {\bf b)} Transmission through the biased $3\times 3$ Au--BDT--Au junction, as compared with the unbiased result, both without \pbc.}
\end{figure}

Finally, we briefly consider calculations of transport through BDT under bias, using the perpendicular-face contacts of Fig.~\ref{fg:unslanted}, and the same basis set as before.
In Fig.~\ref{fg:BDT_bias_profile} we show the potential drop averaged transverse to the transport direction, and observe that the potential is already relatively stable within a few layers of the extended molecule's inner surface. The potential drop is mostly over the thiol end-groups, which may be contrasted with a slightly lower slope of the potential averaged over the core benzene fragment within the extended molecule region, in agreement with the results of Datta \emph{et al.}\cite{Liang2004} and Xue and Ratner.\cite{Xue2002}

We also illustrate the influence of adjusting the location at which the ramp begins further back into the junction. We see that the largest charge accumulation at the interface occurs when the ramp is initiated between the first layers at $-0.5d$ from the innermost Au layer (where $d$ is the interlayer spacing), suggesting that it should begin further back to avoid this. However, we see that as we move the ramp further back, the junction has more difficulty screening the applied potential, reflected in the longer extent of its deviation from $\pm250$mV deeper into the contacts. This implies practical limitations on minimum junction depth, which are likely more relevant in our models without \pbc, given their lower capacity to screen high fields inside the conducting leads.

In Fig.~\ref{fg:AuBDT.bias} we plot the transmission on logarithmic scale, compared with the zero-bias calculation of Fig.~\ref{fg:3x3}. We see that the main effect is the shift in Fermi level and the attenuation of the peaks, as is commonly observed in the presence of an electric field. There do not appear to be further large changes for this relatively small bias, and all prominent features are still clearly recognizable.\\

\section{Conclusions}

A combination of the NEGF approach with a DFT description of the bulk contacts is a flexible, efficient and scalable computational method for transport calculations on realistic geometries of single-molecule devices. A key advantage of our implementation is the ability to break periodicity in 1-3 dimensions within a band-structure code, which allows accurate simulation of the metal contacts for different systems, while allowing us to simulate geometries akin to \emph{e.g.} mechanically controlled break junctions without the need to impose \pbc. 

We have studied one-dimensional chains and found transport behavior that is expected based on simple theoretical considerations, and then extended our scope to three-dimensional junctions involving a series of phenyl-ring molecules from the OPE-family with related gas-phase electronic structures. For benzenedithiol we recover the signature transmission characteristic with \pbc, but if we go beyond this by breaking periodicity we gain a deeper understanding of the more complex transmission-peak structure, as the simple broad HOMO-like peak is separated into smaller peaks that we identify with particular molecular orbitals hybridizing with Au in transport. This may better reflect what happens in experiments with sharp nanocontacts. 

Such characteristic features are seen to evolve in a clear way for the OPE-based molecular wires as well, with the orbitals determining transport near the Fermi level clearly related to the orbital symmetries we identify across the family of molecules. In particular, we see a related cluster of occupied orbitals near $\epsilon_f$ which dominate transport. As we consider progressively longer molecular wires, we find an at first erratic trend, leading finally to convergence in reducing the transport gap between these and the lowest unoccupied-level resonances. Finally, in the low-bias regime, we find that for the simplest phenyl junction the symmetry of coupling ensures that we find no significant spectral shifts, but we do find amplification and attenuation of specific transmission features. 

In looking towards the future, we would note first the utility of the implementation of a gate,\cite{Stadler2008a,Kaasbjerg2008} an end to which work is already underway in our group. Beyond adding a gate as a next step, we envision an extension to a model for the weak-coupling regime using discrete-charge-state Green's functions, along the lines of recent work by Mirjani and Thijssen,\cite{Mirjani2011} for which we anticipate the utility of our underlying charge-constrained DIIS algorithm, here used for convergence acceleration.

\begin{acknowledgments}
We acknowledge financial support by the FOM Foundation within Project-86 and from the EU FP7 program under the ``ELFOS'' grant agreement. This work was also sponsored by the National Computing Facilities Foundation for the use of supercomputer facilities, with financial support from the Netherlands Organization for Scientific Research, NWO. We would also like to thank M. Leijnse for the initial development work in BAND, P. Philipse and A. Yakovlev for their assistance in the development process, and G. Labadze and J.S. Seldenthuis for fruitful discussions.
\end{acknowledgments}

\bibliography{DFT-based-Molecular-Transport-in-Band}


\appendix

\section{Al Chains}\label{Al.Chains}

\begin{figure}
 \includegraphics[width=\columnwidth]{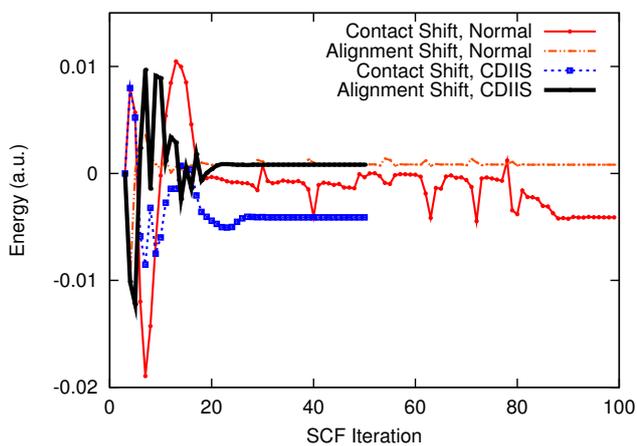}
\caption{Convergence of alignment shifts for the Al chain: $\dphi_{0,1}$ with and without C-DIIS. Note the significantly better performance of the CDIIS procedure.}\label{fg:Al.ShiftConvergence-cdiis}
\end{figure}

Recalling the Li chain in section \ref{Transport1D} we briefly extend the discussion there to a system which \emph{is} physically-realizable: an Al chain. We first treat a homogeneous chain, and then 1D Al contacts to a H$_2$ molecule, with calculations performed using LDA and a DZP-quality basis set on all atoms. The alignment procedure discussed in the main text converges well, and is significantly accelerated by the constrained DIIS extension we discuss in appendix \ref{AlignmentValidation}, which we illustrate in Fig.~\ref{fg:Al.ShiftConvergence-cdiis}.

\begin{figure}
 \includegraphics[width=\columnwidth]{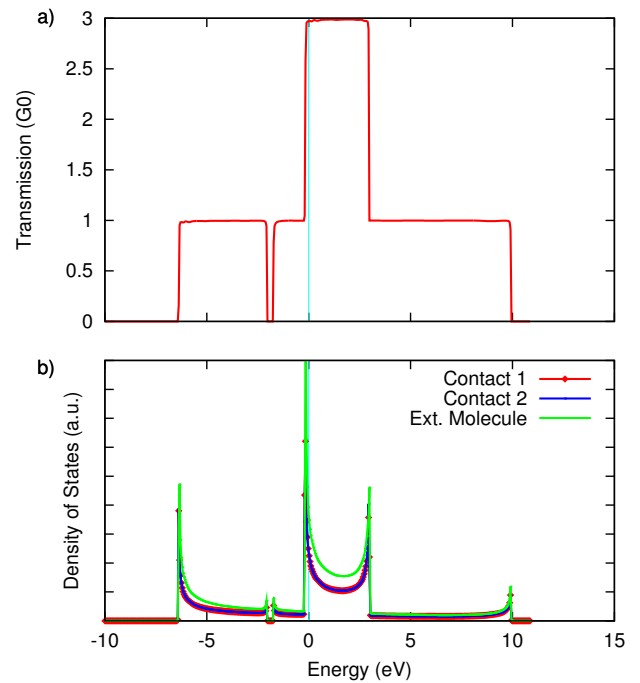}
\caption{(a) Zero-bias transmission through and (b) density of states of the Al chain. Note the position of the Fermi level, determined as offset by 85 meV from the original bulk calculation by our self-consistent alignment procedure. 
}\label{fg:Al.Combination}
\end{figure}

\begin{figure}
 \includegraphics[width=\columnwidth]{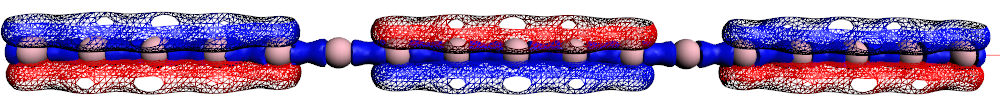}\vspace{.6cm}
\caption{Bulk charge density profile (solid) and HOMO wavefunction (hatched) calculated in the converged alignment configuration for our finite extended Al system, illustrating the 12-atom periodicity of the wavefunction discussed in the text.}\label{fg:Al-Charge-Density}
\end{figure}

The resulting charge density and HOMO wavefunction again compare very well to the bulk 1D calculation, likewise performed with BAND and illustrated in Fig.~\ref{fg:Al-Charge-Density}. This also mirrors what we observed earlier in Fig.~\ref{fg:Li-Charge-Density} for Li chains. 
Considering the projected DOS on the different spatial segments of the calculation in Fig.~\ref{fg:Al.Combination}, we note that the features line up well over the contact regions and the (bulk) extended molecule, with the usual van Hove singularities at the edges of the bands, which are easily identified as the $s-$  and $p-$bands of Al, corresponding to $1G_0$ and $3G_0$ conduction channels respectively.

Turning to the orbitals, in Fig.~\ref{fg:Al-Charge-Density} we observe a signature 12-atom periodicity, analogous to the 4-atom periodicity observed previously in Li. Reconsidering the occupation of the highest level of a chain of fermion sites, we now fill the outer 3-fold degenerate $3p$-orbitals of Al. The argument outlined earlier then leads us to conclude that $\lambda_\text{max} \sim 12a$ (compare equation \eqref{lambda1}):  the wavelength of the highest occupied mode should be precisely 12 lattice spacings, as we indeed find.\\

\begin{figure}
 \includegraphics[width=\columnwidth]{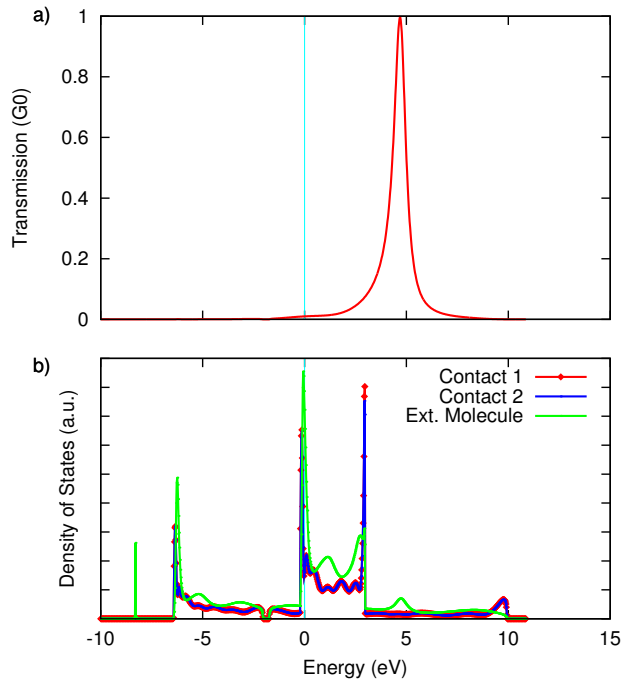}
\caption{(a) Zero-bias transmission through and (b) density of states of the Al--H$_2$--Al junction, together with the $T(\epsilon)$ characteristic of the junction, which is now entirely determined by the H$_2$ molecule in the gap, as compared with Fig.~\ref{fg:Al.Combination}.}\label{fg:AlH2.Combination}
\end{figure}

\begin{figure}
\subfloat[]{\includegraphics[width=.8\columnwidth]{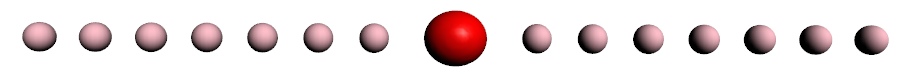}}\\
\subfloat[]{\includegraphics[width=.8\columnwidth]{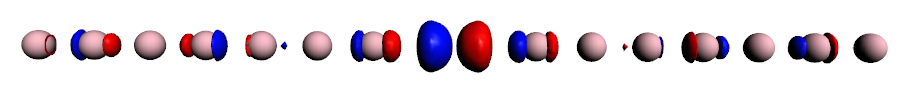}}
\caption{Wavefunctions of the {\bf a}) HOMO and {\bf b}) LUMO orbitals of H$_2$ in zero-bias transport between Al leads.  the LUMO level would be most relevant to transport, corresponding to the resonance in Fig.~\ref{fg:AlH2.Combination}. The HOMO, by contrast, does not hybridize, but shows up as the leftmost peak in that figure, near $-8$ eV.}\label{fg:AlH2.orbitals}
\end{figure}

The converged Al alignment calculation fixes the correct Fermi level for a transport calculation, with the extended-molecule geometry shown in Fig.~\ref{fg:AlH2.orbitals}. The effects of inserting a H$_2$ molecule in the chain of Fig.~\ref{fg:Al.Combination} are reflected in the PDOS for each part of the system in Fig.~\ref{fg:AlH2.Combination}, where we see that they are slightly deformed from their characteristic bulk shapes. In particular, there is a satellite peak corresponding to the $1s^1$ HOMO state on the molecule, and a hump in the extended molecule DOS near 5 eV corresponding to a transmission resonance through the $1s^2$ LUMO state of H$_2$, orbitals illustrated in Fig.~\ref{fg:AlH2.orbitals}.
There, we recognize the origin of the sharpness of the decoupled HOMO peak, in contrast to the broadened LUMO peak which has more strongly hybridized with the leads.

\section{Alignment Validation}\label{AlignmentValidation}

\subsection{Alignment Examples}\label{AlignmentTests}

\begin{figure}
 \includegraphics[width=\columnwidth]{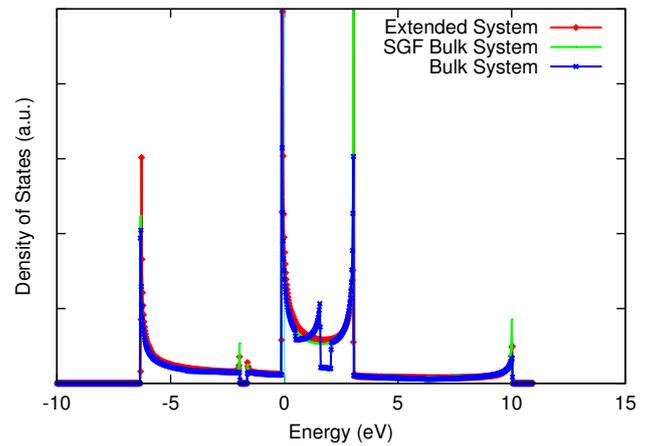}
\caption{Alignment PDOS for infinite Al chain contacts (periodic along transport direction), compared with the bulk calculation of the same chain contacts. LDA-level DFT calculations, with a double-$\zeta$ basis used; $k$-space calculations beyond the $\Gamma$-point approximation are necessary to reproduce the feature near 2 eV in the bulk system.}\label{fg:Al.align_dos}
\end{figure}

We briefly discuss the validation of the alignment procedure in  the context of this constrained DIIS algorithm.
For the case of Al chains, we previously illustrated the result of alignment in Fig.~\ref{fg:Al-Charge-Density}, and the density of states projected onto the different segments of the calculation is shown in Fig.~\ref{fg:Al.align_dos}. We conclude from these that the calculations reproduce the bulk results well, as further evidenced by Fig.~\ref{fg:Al.Combination}. The major features line up over the contact regions and the extended molecule, and the calculations reproduce easily identified $s-$ and $p-$bands corresponding to $1G_0$ and $3G_0$ conduction channels respectively.\\

\begin{figure}
 \includegraphics[width=\columnwidth]{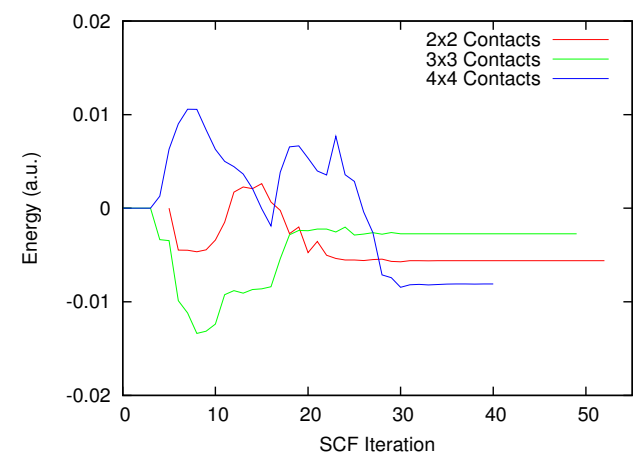}
 \includegraphics[width=\columnwidth]{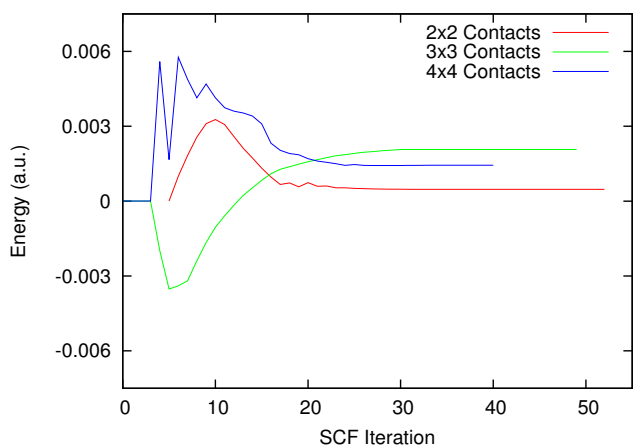}
\caption{Alignment shift convergence for Au contacts without periodic boundary conditions, for $2\times 2$, $3\times 3$ and $4\times 4$ contacts. {\bf a)} the contact shift (zero of the potential) $\dphi_0$, and {\bf b)} the alignment shift $\dphi_1$. }\label{fg:Au.ShiftConvergence1}
\end{figure}

\begin{figure}
 \includegraphics[width=\columnwidth]{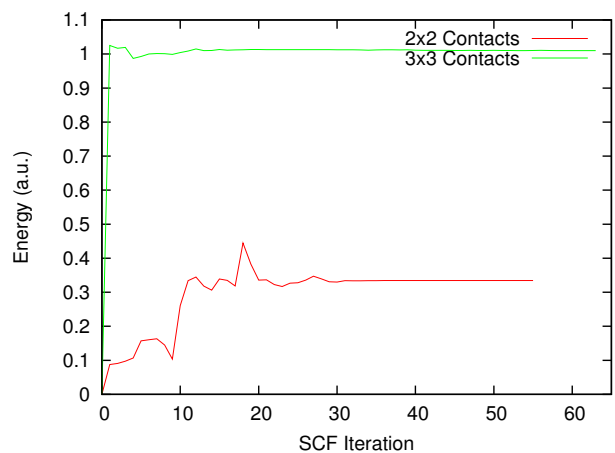}
 \includegraphics[width=\columnwidth]{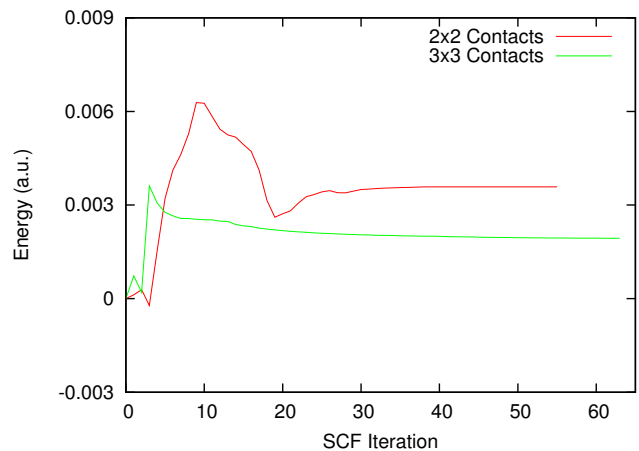}
\caption{Alignment shift convergence for Au contacts with \pbc, for $2\times 2$ and $3\times 3$ contacts. {\bf a)} the contact shift (zero of the potential) $\dphi_0$, which is much larger in magnitude for the case where \pbc\; are applied in contrast to {\bf b)} the alignment shift $\dphi_1$, of similar magnitude those in Fig.~\ref{fg:Au.ShiftConvergence1}.}\label{fg:Au.ShiftConvergence2}
\end{figure}

\begin{figure}
 \includegraphics[width=\columnwidth]{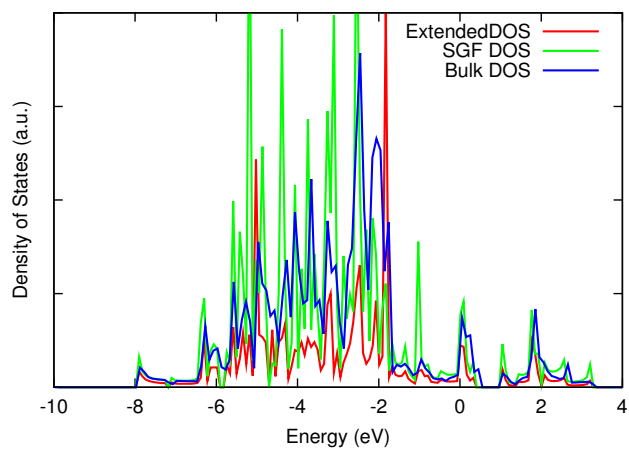}
\caption{Converged PDOS for $2\times 2$ Au contacts and ``bulk'' extended molecule, as compared with the bulk DOS for Au with the same basis-set; agreement is also typical for larger contacts.
}
\label{fg:Au.PDOS.Compares}
\end{figure}

We next turn our consideration to the case of ``bulky'' Au contacts, also discussed in the main text. In Fig.~\ref{fg:Au.geoms}, we showed representative geometries, which were all FCC stacked with a (111) face perpendicular to the transport direction. Here we compare alignment of different sized transverse surfaces, and for a $3\times 3$--surface the results both with and without {\pbc}.

In the systems without \pbc, 
we find comparable offsets in Fig.~\ref{fg:Au.ShiftConvergence1}. In Fig.~\ref{fg:Au.ShiftConvergence2} by contrast, we immediately see that the shifts necessary to align the ``bulk'' extended-molecule structure with the bulk contacts are quite large, while the fine-tuning for charge neutrality is comparable to the cases without \pbc. We emphasize that this is an algorithmically-determined shift, and that as we illustrate in Fig.~\ref{fg:Au.PDOS.Compares}, the PDOS indicate that the structure is correctly converged.

\section{Constrained DIIS Method}\label{DIIS}

In normal Pulay DIIS,\cite{Pulay1980} the SCF process is cast as a linear(-izable) mapping $F(V)$ of some vector $V$, which in DFT are the Fock operator and the density of the system (in discretized real space) respectively. Normally one would simply have:
\begin{align}
V\itr{n+1} = F\left(V\itr{n}\right)
\end{align}
for the $n+1^\text{st}$ iteration, and iterate until convergence. Pulay's insight was to seek the optimal combination of a history of previous iterations (an iterative subspace):
\begin{align}
V\itr{n} &= \sum_{i=0}^{n-1} c_i V\itr{i}
\end{align}
subject to $\sum_i c_i = 1$ for normalization of the result. Let the iterates $V\itr{n}$ denote the input to the Fock operator, and denote the output by $V_*\itr{n} = F\left(V\itr{n}\right)$. Then we also have a history
\begin{align*}
V_*\itr{n} &= \sum_{i=0}^{n-1} c_i V_*\itr{i}\;,
\end{align*}
trivially. To quantify the quality of this linear combination, a norm is introduced on the error:
\begin{align}
V_*\itr{n}-V\itr{n} &= \sum_i c_i\left(V_*\itr{i}-V\itr{i}\right) \quad\text{such that}\\
\nonumber \|V_*\itr{n}-V\itr{n} \|&=\sum_i\sum_j c_ic_j\langle V_*\itr{i}-V\itr{i} | V_*\itr{j}-V\itr{j}\rangle\\
&= \bm{c}^\top B\bm{c}
\end{align}
where $\bm{c}=\left(c_0,\ldots,c_{n-1}\right)^\top$ and $B$ has elements $b_{ij} = \langle V_*\itr{i}-V\itr{i} | V_*\itr{j}-V\itr{j}\rangle$ formed by an inner product.
This can be recast as a cost function including the constraint:
\begin{align}
J &= \bm{c}^\top B\bm{c} -\lambda \bm{c}^\top\mathds{1}\;,
\end{align}
minimized by:
\begin{align}
\nabla J &= 0 \ni B\bm{c} = \lambda\mathds{1}\;.
\end{align}
Solving this is precisely direct inversion of the iterative subspace, and the Lagrange multiplier is $\lambda = \displaystyle \frac{1}{\sum_i c_i}$.\\

During and alignment run we extend this approach as follows. Our goal is to allow the SCF cycle, using the DIIS algorithm, to become aware of the constraint that at convergence,
\begin{align*}
\tr{\rho S}_\text{\tiny EM} \rightarrow Q_\text{\tiny EM}
\end{align*}
should hold, with $Q_\text{\tiny EM}$ the correct bulk (valence) charge on the EM composed of bulk material. Observing that this is already in the same units of charge as the relevant vector $V$, the DIIS approach can be extended, defining the excess charge on the extended molecule $Q_x \equiv \tr{\rho S}_\text{\tiny EM} -Q_\text{\tiny EM}$, as follows:
\begin{align}
V   \rightarrow \left( V ,\; Q_x\, \right)
\end{align}
\emph{i.e.} extending the vector by a single scalar, and
\begin{align}
b_{ij} \rightarrow b_{ij} + \left(Q_{x*}\itr{i} - Q_x\itr{i}\right)\left(Q_{x*}\itr{j} - Q_x\itr{j}\right)\;,
\end{align}
where $B$ preserves its meaning as the matrix of error norms or correlations. Note that DIIS convergence implies that the vector (\emph{i.e.} the density) becomes stationary, and likewise for the excess charge. 

We can also choose to use $b_{ij} + Q_x\itr{i} Q_x\itr{j}$ instead, which is the formally more desirable condition that the excess charge go to $0$ directly, but then we depart somewhat from the spirit of the Pulay approach.

\section{Transmission Peak Decompositions}\label{decompositions}

In order to understand the composition of the peaks present in the transmission through a junction, we 
represent the contribution of the fragments orbitals to the transmission peaks
as follows.

From a DFT calculation of the fragment in BAND we store the eigensystem $\{\epsilon_j,\;|\phi_j\rangle\}$ of the fragment. 
We then perform the transport calculation in band, obtaining the transmission from
\[ T(\epsilon)= \trr{\; \Gamma_1(\epsilon) G(\epsilon) \Gamma_2(\epsilon) G^{\dagger}(\epsilon) \;} \]
which contains the Green's function
\[ G(\epsilon) = \biggl( \epsilon\,S - H -\left(\Sigma_1(\epsilon)+\Sigma_2(\epsilon)\right)\biggr)^{-1}\;. \]
At the end of the SCF calculation, in addition to calculating transport properties, BAND also diagonalizes the (aligned) Fock matrix $H$ to obtain a discrete set of transporting levels
$\{\varepsilon_i,\;|\psi_i\rangle\}$.

We use the fragment-calculation functionality of the ADF package to project the subset of molecular levels of interest (HOMO, LUMO, \emph{etc}.) onto the full set of transporting levels, and obtain a table of the projections $|\langle \psi_i|\phi_j \rangle|$, which are the ``peak decompositions''\footnote{Strictly speaking, because we do not project onto the density matrix obtained from the Green's function, it is a fragment projection, rather than a true peak decomposition, since the peaks are obtained from the full $G(\epsilon)$.} we referred to in the main text. Each component $0\leq |\langle \psi_i|\phi_j| \rangle \leq 1$ tells us to what extent the transporting level is composed of the $|\phi_j \rangle$, which we then visualize in Fig.~\ref{fg:bdt.peaks} as a stacked bar chart. The bar chart is overlaid onto the transmission, with each bar centered at the corresponding $\varepsilon_i$. 

We remark that the correspondence between this discrete set of decompositions and the transmission $T(\epsilon)$ is not exactly one-to-one, given that the spectrum $\varepsilon_i$ is that of the Fock matrix $H$, while the running variable $\epsilon$ in the transmission corresponds to the spectrum of $G(\epsilon)$, which includes the effect of $\Sigma_1(\epsilon)$ and $\Sigma_2(\epsilon)$. It is nonetheless a rather good correspondence, as is verified by 
considering the levels $|\psi_i\rangle$ nearest the $\epsilon$ of a given peak in the transmission.

\end{document}